**Author for correspondence:**
Patricia Schady
e-mail: pschady@mpe.mpg.de

# Gamma-ray bursts and their use as cosmic probes

Patricia Schady

Max-Planck-Institut für Extraterrestrische Physik, Giessenbachstraße, 85748 Garching, Germany

PS, 0000-0002-1214-770X

Since the launch of the highly successful and ongoing *Swift* mission, the field of gamma-ray bursts (GRBs) has undergone a revolution. The arcsecond GRB localizations available within just a few minutes of the GRB alert has signified the continual sampling of the GRB evolution through the prompt to afterglow phases revealing unexpected flaring and plateau phases, the first detection of a kilonova coincident with a short GRB, and the identification of samples of low-luminosity, ultra-long and highly dust-extinguished GRBs. The increased numbers of GRB afterglows, GRB-supernova detections, redshifts and host galaxy associations has greatly improved our understanding of what produces and powers these immense, cosmological explosions. Nevertheless, more high-quality data often also reveal greater complexity. In this review, I summarize some of the milestones made in GRB research during the *Swift* era, and how previous widely accepted theoretical models have had to adapt to accommodate the new wealth of observational data.

## 1. Introduction

During the first couple of decades after the first reported detection of a gamma-ray burst (GRB) by the USA military Vela satellites [1], the origin of these vast explosions perplexed theoreticians and observers alike. Early theories ranged from nearby white dwarfs to extraterrestrial activity, with many finding it inconceivable that such explosions could arise from extragalactic environments, given that the implied isotropic-equivalent energies would reach up to a few tenths of a solar mass, all released in just a few tens of seconds. Nevertheless, very soon after the launch of the Compton Gamma Ray Observatory (CGRO) in 1991, the on-board Burst and Transient Source Experiment (BATSE) instrument showed that GRBs were isotropically distributed [2,3]; clear evidence that GRBs were indeed extragalactic.

The next momentous leap in the investigation of GRBs came with the first detection of an X-ray [4] and optical counterpart [5]

to GRB 970228,[1] which was largely made possible by the comparatively accurate (several arcminutes), and early position (few hours after the GRB) available with the newly launched Italian–Dutch BeppoSAX satellite [6]. Although previous searches for the expected counterparts of GRBs had been carried out [7,8], they were typically too late after the GRB, and too shallow to detect the rapidly fading afterglow. The sub-arcsecond optical localization of GRB 970228 placed it within the outskirts of an underlying faint galaxy [5,9], further strengthening the case that GRBs had an extragalactic origin. However, the final crunch came a few months later, with the eventual spectral observation of the optical counterpart of GRB 970508 [10] and the subsequent redshift determination of $z = 0.835$ [11], thus confirming the extragalactic nature of GRBs. The detection of the GRB 'afterglow' signified much more than just a distance measure. The much longer lived, lower energy emission provided the opportunity to study the light curve and spectral evolution of the GRB; it pinpointed the GRB position to sub-arcsecond accuracy, thus enabling the host galaxy to be identified, and it offered the chance to study the absorption imprint left on the afterglow by intervening material within the host galaxy.

With the launch in 2004 of the GRB-dedicated NASA *Swift* mission [12] came the next leap in our understanding of GRBs. In addition to the GRB alert telescope (BAT; [13]), *Swift* has an X-ray telescope (XRT; [14]) and an ultraviolet and optical telescope (UVOT; [15]). Its very rapid slewing mechanism has drastically increased the detection rate of GRB afterglows, and decreased the typical delay times between the high energy prompt emission and longer wavelength afterglow emission from hours to minutes. These data have opened up a new parameter space, and with that there have been many surprises from the perspective of predictions that turned out to be unfounded, and newly observed features that were unexpected based on standard theoretical models. These include very variable X-ray afterglow light curves, long-lived afterglow plateaus, new populations of intrinsically low-luminosity and heavily dust extinguished GRBs, and a subsequent significant increase in the variation in host galaxy properties. These enlightening data have engulfed the GRB community with an unprecedented amount of information on the multi-wavelength spectral and temporal properties of GRBs and on the environmental conditions that they trace, which have shaken up many long-standing progenitor models, and theories on the prompt and afterglow emission mechanisms that were largely taken as truths.

In this review, I will focus on the developments in GRB research during the last decade of *Swift* GRB observations, primarily with respect to the optical and X-ray afterglow and host galaxy observations. For information on the pre-*Swift* state of the field and the details of the well-established GRB 'collapsar' and fireball model, I refer the interested reader to some of the excellent reviews written on these topics, such as [16–18] and papers therein. This review article is structured as follows. In §2, I give a brief description on the various subsets of GRBs identified during the *Swift* era (§2.1), and the implications for the diversity of GRB progenitors. I use this section to discuss the properties and evolution of the GRB collimated outflow and how this affects observations (§2.2), the evidence for long-lived energy injection that has obliged us to reassess the GRB central engine model (§2.3), and the progress that has been achieved in understanding the progenitors of short and long GRBs through the GRB-supernova (§2.4) and GRB-kilonova (§2.5) connection. I dedicate the second half of this review to the long-duration variety of GRBs, which are linked to the stellar evolution of massive stars and thus act as powerful probes of star formation in the distant Universe. I first provide a summary in §3 of our current understanding on the host galaxies of long GRBs from galaxy emission data, and discuss the implications for the use of long GRBs as probes of star formation. In §4, I then discuss the unique vantage point provided by long GRBs on the properties of the interstellar medium (ISM) within distant, star-forming galaxies from the absorption imprint left on the GRB broadband afterglow. Finally, in §5, I conclude with a brief outlook of the future for GRB science in the era of multi-messenger astronomy, and the prospects for understanding many of the outstanding issues within progenitor models.

# 2. Gamma-ray burst classification schemes and progenitor models

## 2.1. Short, long and ultra-long duration gamma-ray bursts

The GRB prompt gamma-ray emission is highly variable, giving rise to multi-peaked light curves with a range of delays between each new pulse of radiation. These pulses frequently overlap such that they are difficult to isolate, but they can also be separated by long gaps in the $\gamma$-ray emission that lasts longer than the duration of the pulses themselves. For this reason, the $T_{90}$ parameter was devised to quantify the duration of GRB prompt emission light curves, and which is defined as the (observer frame) time interval that contains 90% of the GRB $\gamma$-ray fluence. Using a sample of 222 GRBs from the first BATSE catalogue, [19] found a clear bimodal distribution in the $T_{90}$ of GRBs separated at roughly 2 s, evidence of which was already present in the early KONUS GRB catalogue [20]. The BATSE data also showed distinct differences in the hardness ratio of short ($T_{90} \lesssim 2$ s) and long ($T_{90} \gtrsim 2$ s) GRBs, which led to the 'short-hard' and 'long-soft' GRB classification scheme. The significance of a bimodal distribution in $T_{90}$ and in the hardness ratio was greatly strengthened with the full 2704 GRBs detected with BATSE and later GRB missions, providing strong evidence for two GRB progenitor channels.

[1]GRBs are named by date of detection, following the syntax GRB YYMMDD. Prior to 2010, if more than a single GRB was detected on the same day, the first GRB then took the suffix 'A', the second GRB took the suffix 'B' and so on. Owing to confusion in the literature with the 'A' suffix not always being added to the GRB name when a second GRB occurred on the same day, in 2010 the convention was changed so that the first GRB detected on any given day would always carry the suffix 'A'.

There is significant overlap in the distribution of short and long GRBs in the $T_{90}$-hardness ratio parameter space, and thus a robust classification scheme should naturally rely on physical properties. This is exemplified by the population of short GRBs with extended emission, which are characterized by an initial short, spectrally hard $\gamma$-ray pulse followed by much dimmer and softer emission lasting for tens of seconds [21]. When categorizing GRBs by just the $T_{90}$ and spectral hardness of the prompt emission, an extended emission GRB could be classified as both long and short depending on the sensitivity and energy range of the GRB alert instrument. Whereas long GRBs arise from the core collapse of a massive star, referred to as the 'collapsar' model [22,23], short GRBs are thought to form from the binary merger of two neutron stars (NS) or an NS–black hole system, and a working definition of short and long GRBs that takes into account these different progenitor channels is preferable. Prior to the launch of *Swift*, there had been no afterglow detection, and thus no arcsecond localization, of a short GRB. However, with the rapid *Swift* follow-up observations came the first afterglow detections and host galaxy identifications, providing greater discriminating power. Deep follow-up observations of well-localized short GRBs have found no emerging supernova (SN) down to deep limits, contrary to long GRBs (see §2.4), and the associated host galaxies are often elliptical galaxies with no ongoing star formation [24–26], providing supporting evidence that short GRBs are not related to massive star formation.

A more recent class of 'ultra-long' GRBs has emerged, largely prompted by the extremely long-lived 'Christmas-day burst', GRB 101225A [27,28], which had a $\gamma$-ray emission light curve that lasted for more than 7000 s [28,29]. A handful of previous GRBs detected with a number of instruments (BATSE, Konus-*Wind*, BeppoSAX and *Swift*) had comparatively long $\gamma$-ray light curves [30–33], but these are rare events, and it remains unclear whether ultra-long GRBs represent a distinct population of bursts that have different progenitors to classical 'normal' long GRBs [29,34,35], or whether they are an extension of the same population, making up the high-end tail of the $T_{90}$ distribution [33,36]. As eluded to above, defining a class of GRBs by $T_{90}$ alone is ambiguous, as the measured $T_{90}$ varies with energy range. Furthermore, there are selection effects in the detection of very long-duration, low-luminosity GRBs, which may affect the sampling of the long tail of the $T_{90}$ distribution [29,36]. In light of this, Levan *et al*. [29] proposed a definition of ultra-long GRBs that includes multi-wavelength criteria, consisting of very long-duration prompt emission light curves (observed for more than 10 000 s at $\gamma$ and X-ray wavelengths), short-scale variations during a luminous X-ray plateau phase (see §2.3), and very rapid decay rates ($\alpha > 3$ where $F_t \propto t^{-\alpha}$) at the end of the X-ray plateau, as expected from the sudden cessation of the central engine. Using these criteria, Levan *et al*. [29] identified four *Swift* GRBs which they defined as ultra-long,[2] and as such, as having a different emission mechanism to classical long GRBs. They identified several more possible ultra-long GRBs that satisfied just part of their criteria. Several progenitor channels have been considered to power the outburst of ultra-long GRBs for such an extended period of time [37], including the tidal disruption event (TDE) of a star by the galaxy central black hole [38], the core collapse of a low-metallicity blue supergiant into a black hole (BSG; [29,34]), and the core collapse-induced formation of a highly magnetized NS, or 'magnetar' central engine rather than a black hole-accretion disc-powered event [39]. The first spectroscopic detection of an SN coincident with an ultra-long GRB shows tantalizing evidence that ultra-long GRBs and the new superluminous class of supernovae (SLSNe) are related [39] (see §2.4.2).

## 2.2. The effect of viewing angle and jet opening angle

A common ingredient in all GRB models of the past two decades (for short and long GRBs) is that the initial high energy emission, and the longer-lived, very broad wavelength afterglow (from X-ray to radio frequencies) are released within narrow jets that are powered by a central engine [40–43]. An expanding

[2]pre-*Swift* GRBs are typically excluded from the selection criteria by definition due to the lack of early-time X-ray data prior to the launch of *Swift*.

fireball accelerates the ejecta to relativistic velocities (bulk Lorentz factors $\Gamma = 100$–1000) [16], and a combination of synchrotron and inverse Compton radiation from electrons accelerated within internal shocks [44,45], and possibly also thermal electrons released from the photosphere [46–48], contribute to the highly erratic $\gamma$-ray light curve. The afterglow emission is somewhat simpler, with a suite of evidence supporting an external shock model. As the GRB jet is slowed down by the surrounding circumburst medium, a forward and a reverse shock form. Accelerated electrons within the shocked region then cool through synchrotron emission, giving rise to the broadband afterglow.

The release of emission through jets greatly reduces the energy reservoir required to power the GRB by a factor of several, substantially alleviating the conditions for progenitor models. However, the consequence is that the intrinsic, jet-corrected energy of a GRB and the true GRB rate are poorly constrained, and it is often difficult to differentiate between geometric and dynamical effects. For example, some very soft GRBs, also known as X-ray flashes [49] could be the result of GRBs viewed off-axis, whereas a more recently defined class of low luminosity GRBs may be the result of large jet opening angles [50], (see [51] for a review on the various GRB progenitor groups).

The corrections that need to be applied to go from observed to intrinsic properties require knowledge of the jet opening angle and the observer viewing angle relative to the jet axis, both of which are very challenging to measure. A predicted observational signature of jet emission is the presence of achromatic 'jet-breaks' in the afterglow light curve across all observable frequencies, which arises when the relativistic emission from the entire surface of the jet is observable and when the jet begins to spread sideways. At high bulk Lorentz factors the emission is narrowly beamed such that emission from only a small fraction of the ejecta is detected at a given time. As the ejecta decelerates and the Lorentz factor decreases, the beaming angle becomes larger, thus bringing a larger fraction of the emitting region into view until the observer sees emission from the entire jet. Continuous deceleration of the ejecta and subsequently larger beaming angles no longer increase the observable emitting region, producing a telltale 'jet break' in the light curve across the entire afterglow spectrum [52]. On a similar timescale, the deceleration of the jet may also cause the jet to start spreading sideways, further reducing the observed GRB emission [43,52,53]. The sharpness of this jet break and the change in the afterglow decay rate depends on how long the jet can remain collimated [54], and on the jet radial density profile and energy distribution [54–56]. The time of the jet break is instead related to the jet opening angle, the bulk Lorentz factor and the density of the circumburst medium.

Prior to the launch of *Swift*, simultaneous breaks in the optical and near-infrared (NIR) afterglow light curves were frequently observed and were interpreted as jet breaks. The implied GRB beam-corrected kinetic energy presented tantalizing evidence that at least long GRBs were powered by a standard energy reservoir, and that the large variation in the isotropic-equivalent energy observed in GRBs was the result of a large range in jet opening angles [57,58]. Such a possibility had exciting implications for the use of GRBs as standardizable candles. Nevertheless, since the launch of *Swift*, the orders of magnitude improved temporal and spectral coverage of GRB afterglows, in particular at X-ray energies with the very rapid *Swift*/XRT observations, have revealed much more complex afterglow light curves than implied by pre-*Swift* data, leaving many pre-*Swift* jet-break identifications rather ambiguous. Within the first few hours after the prompt emission, where afterglow observations had been rare prior to *Swift*, it is now common to observe large flares, unexpected plateaus and chromatic breaks that are not consistent with jet breaks or any synchrotron spectral breaks predicted by standard models. On the other hand, the number of detected achromatic breaks (especially in the optical and X-ray wavelength range) are now relatively small. As a result of these new revelations, previously identified 'jet-breaks' have now been put in doubt, and the notion of a standard GRB energy reservoir has been ruled out.

The absence of clear jet-break signatures in the large fraction of well-observed GRB afterglows has been interpreted as the over-simplification of theoretical models that assumed homogeneous jets with sharp edges, and which did not consider the complex evolution of the afterglow synchrotron emission spectrum. More complex models now include structured jets or multiple embedded jets with different opening angles that produce several chromatic jet breaks [59], or much smoother breaks that may be difficult to detect in the absence of a wide temporal baseline [60]. Moreover, quite recent sophisticated numerical simulations that follow the evolution of the jet [61,62] imply that jets can keep their structure for longer than previously thought, thus delaying the onset of any jet breaks produced by sideways spreading. The increased complexity of the jet dynamics, and the implications in some models that jet breaks do not arise for tens of days after the GRB prompt phase greatly reduces the chance to measure the jet opening angle, thus increasing the uncertainties on the GRB energetics and the intrinsic rate of the short- and long-duration class [63].

## 2.3. The gamma-ray burst central engine

Although the fundamental predictions of the collapsar model and afterglow synchrotron emission properties have withstood the test of time remarkably well, there are many examples where greater complexity in the models is needed to reproduce the GRB afterglow observations. Such relatively common but unanticipated features include very luminous X-ray flares occurring up to a few $10^4$ s after the GRB [64–67], achromatic breaks in the X-ray and optical afterglow light curves, and extended plateau phases that last for a few hours during the early afterglow evolution [68,69]. To account for these unexpected afterglow properties, the fireball model has been supplemented by more complex jet structures that give rise to distinct X-ray and optical emission regions [70–72], evolving micro-physical parameters within the forward shock region, such as a time dependence on the fraction of energy contained within the accelerated electrons and magnetic field [73], and long-lived energy injection [74,75]. Whereas it is not so surprising, retrospectively, to find that the jet dynamics and the evolution of the shocked region are more complex than the simplified prescriptions employed in the standard fireball [76], the evidence for extended central-engine activity tens of thousands of seconds after the GRB was largely unanticipated.

Around 40% of X-ray afterglows observed with *Swift* XRT have flares, and the shape [64,65,77] and spectral [66,78] similarities of these flares with those flares observed during the prompt emission suggest a common origin and imply ongoing central-engine activity out to approximately 1000 s, and maybe even as late as $10^4$–$10^5$ s [67] after the GRB. More compelling are the extended plateau phases present in around a third of X-ray afterglow light curves that last for a few $10^4$ s [68,69,79]. From the standard external shock model, the pre-jet break X-ray afterglow is expected to decay with an index $\alpha \gtrsim 0.8$ where $F \propto t^{-\alpha}$. However, the observed plateau phases have decay indices $\alpha = 0.1$–$0.3$. Such slow evolution of the light curve may arise in a two-component jet or jet-cocoon model [70], or more probably it is produced by a continuous source of energy injection lasting the approximately $10^4$ s duration of the plateau phase.

A property of the plateau phase that can provide further insight on its origin is an anti-correlation between the X-ray luminosity at the end of the plateau phase and the rest-frame plateau end time in long GRBs (the LTX correlation) [80–82]. Such a correlation also appears to exist in the X-ray afterglows of short GRBs, although the normalization is offset such that, for the same plateau rest-frame end time, short GRBs are less X-ray luminous [83,84]. An equivalent anti-correlation between the *optical* luminosity and the plateau end-time (LTO correlation) has also been detected in a subset of long GRB optical afterglows with evidence of a plateau phase [85,86], and a large number of short GRBs with good coverage of the optical afterglow also show very flat light curves at early times [87–89]. However, the significance of an LTO anti-correlation in short GRBs has not yet been explored (although see [90]). An important consequence of these afterglow relations is that any model put forward to explain the plateau phase must also be able to account for the LTX and LTO correlations.

Although multiple emission components stemming from a two-component jet, for example, may be able to account for a plateau phase in the GRB early-time afterglow, they cannot produce the observed LTX and LTO correlations [91]. The correlations thus imply that there must be a continual source of energy injection either from a long-lived central engine, or from slower shells of ejecta that catch up and re-energize the forward and reverse shock at later times [91]. The common detection of short X-ray flares for tens of thousands of seconds after the GRB is also indicative of a persistent source of energy injection [67,68,74]. Such long-lived energy injection from a persistent black-hole-powered central engine requires a significant mass of material to accrete onto the black hole on similar timescales to the duration of the plateau phase. This could occur if a large accretion disc formed at the time of core collapse (approx. $1\,M_\odot$), which would require a very low disc viscosity ($\alpha < 10^{-2}$), or if fall-back material continuously replenished the accretion disc at a similar rate as material was accreted onto the black hole [92–94].

An alternative scenario that has received renewed attention is that rather than a black hole, the central engine is a rapidly spinning, highly magnetized NS, or magnetar, which could form through the accretion-induced collapse of a white dwarf, the collapse of a massive star, or the merger of two NSs [95–99]. A newly formed magnetar would naturally produce a plateau phase via dipole radiation, which would continue until it reached the dipole spin-down timescale [99–102]. One of the reasons why a black hole plus accretion disc central engine has gained more traction in the past is because there is more flexibility in how much energy can be extracted from such a system. In contrast with this, there is a limit on how much energy can be extracted from a magnetar, and very high efficiencies have had to be imposed in order for the magnetar model to have sufficient luminosity to satisfy the observations [84]. For nearby long GRBs where an accompanying SN has been well observed, in general it does not seem possible to extract sufficient energy from a magnetar central engine to power both the GRB and accompanying SN [103].

## 2.4. The long gamma-ray burst–supernova connection

Very early on in the field of GRB research, and before the first afterglow detection, a connection between the core collapse of massive stars was already being made [22], and the smoking gun came with the association of the long GRB 980425 at $z = 0.0085$[3] with the broad line Type Ic (Ic-BL) supernova SN 1998bw [104], a year after the first afterglow detection. However, the orders of magnitude less luminous $\gamma$-ray emission associated with GRB 980425 and the lack of an optical afterglow detection led to some speculation on whether this event was representative of the more standard luminous class of long GRBs. Eventually, the connection between luminous long GRBs and Type Ic-BL SNe was sealed with the detection of SN 2003dh temporally and spatially coincident with GRB 030329 [105,106] at $z = 0.1685$, which had an isotropic-equivalent $\gamma$-ray energy and optical afterglow light curve far more comparable to other cosmological GRBs.

The launch of *Swift* has increased the sample of spectrally confirmed GRB–SN associations by a factor of a few. At larger redshifts ($z \gtrsim 0.5$), where spectroscopic verification of an emerging SN is unattainable, there is also an increasing sample of GRBs showing evidence of SN emission in the form of bumps in the afterglow light curve appearing 10–30 days (observer frame) after the GRB. Most long GRBs have isotropic-equivalent $\gamma$-ray energies $E_{\gamma,\rm iso} \approx 10^{51} - 10^{52}$ erg, whereas the overall population of spectrally confirmed GRB-SNe have $E_{\gamma,\rm iso} < 10^{49}$ erg [104,107,108]. This is probably a result of selection effects whereby the more common (per unit volume) low-luminosity GRBs [109] are not detected at high redshift, whereas luminous, long GRBs have a higher detection rate at higher redshift, where the available volumetric area is larger.

Intriguingly, the SNe associated with low luminosity and with cosmological GRBs all have very similar spectra, and they have a fairly narrow spread in peak luminosities, suggesting that the progenitors of low- and high-luminosity GRBs are similar. Whereas GRBs with associated SNe have isotropic luminosities that span six orders of magnitude, the accompanying SNe only span two orders of magnitude in absolute peak magnitude [50]. It is therefore some property of the GRB emission mechanism other than the progenitor itself that generates the large differences in the isotropic energy. One such important property may be the rotational velocity of the stellar core at the time of gravitational collapse, which will shape the dynamics of the ensuing jet. This is the basis for the suggestion that the jet formed in low-luminosity GRBs 'fails' to break out of the stellar envelope, and the GRB is instead powered by less-energetic and isotropic shock break-out emission [50,110]. One might then expect to see various correlations between the energy of the GRB and afterglow and certain environmental properties, such as metallicity, which strongly affects the stellar mass loss and thus rotational velocity during the lifetime of the progenitor star. Although there is strong evidence that the long GRB population as a whole has a preference for lower metallicity environments [111–114], there is no evident correlation between the GRB and environmental properties [115]. Nevertheless, it remains unclear what the mechanisms involved in launching the jet are and what the main contributing factors are that define its structure. The lack of an obvious relation between the GRB energetics and environmental properties may thus reflect the complexity in the formation of the jets, and the general difficulty in measuring the environmental conditions in the immediate vicinity of the GRB (§2.2).

Somewhat in discord with the picture that is emerging of long GRB–SN, there are at least two cases of nearby long GRBs (GRB 060505 and GRB 060614) where, despite extensive follow-up campaigns, no associated SNe were detected down to deep limits. Any accompanying SN must have been a hundred times fainter than SN 1998bw [116–118]. Another recent addition to this population of supernovaless long GRBs is that of GRB 111005A at $z = 0.013$, for which deep *Spitzer* observations imply an upper limit on an accompanying SN that is 20 times less luminous than any previously detected GRB–SN [119]. These few cases of nearby long GRBs with no associated SN emission may signify the existence of more exotic long GRB formation mechanisms, or they may be examples of 'failed' SNe [120]. Nevertheless, the more than three orders of magnitude difference in the isotropic energy of all three supernovaless events (from $E_{\gamma,\rm iso} \sim 2 \times 10^{47}$ erg for GRB 111005A up to $E_{\gamma,\rm iso} \sim 9 \times 10^{50}$ for GRB 060614) make the connection between these three GRBs unclear. In all three cases the GRB redshift originates from the association of the GRB with a galaxy spatially coincident with the afterglow position. Although unlikely, in each case the chance alignment of a foreground galaxy along the GRB line of sight can therefore not be ruled out

[3]GRB 980425 continues to be the nearest GRB (long and short) detected.

(e.g. [121]). Future, concrete examples of such events with absorption-based spectroscopic redshifts from the afterglow are therefore preferable to explore further possible progenitor models.

Below, I focus on two exceptional long GRBs where the detection of an accompanying SN signified a benchmark in our progress on understanding the GRB–SN connection and on the relation between different subcategories of long GRBs. For a more in-depth overview of the GRB–SN connection, I refer the interested reader to [50].

### 2.4.1. High-luminosity gamma-ray bursts and their SNe: the case of GRB 130427A/SN 2013cq

Much of what we know about GRB progenitors and their emission mechanisms comes from a few, very well-observed cases, such as GRB 030329/SN 2003dh, and another such example was the detection of the exceptional event, GRB 130427A [122–125], at a redshift $z = 0.3399$. GRB 130427A had an isotropic energy $E_{\gamma,\mathrm{iso}} \sim 8 \times 10^{53}$ erg [123,126,127], making it one of the most energetic GRBs ever detected and the most energetic GRB at $z < 0.5$ by almost two orders of magnitude. Moreover, it had a clearly detected associated SN, providing an important link between nearby GRB-SNe and very distant and energetic GRBs. The immense luminosity of GRB 130427A and its proximity enabled its evolution to be monitored in exceptional detail over 16 decades in wavelength space, and it was still visible in X-rays more than 3 years after the GRB explosion [125]. In contrast with the GRB properties, the accompanying SN 2013cq contained a comparable energy release to previous GRB-SNe [128], re-affirming previous indications that nearby low-luminosity GRBs and high-redshift GRBs have a common progenitor.

The high-quality dataset available of the afterglow and SN accompanying GRB 130427A provides a rigorous test of standard GRB afterglow theory, which was developed around two decades ago [129], and is based on approximations of the acceleration processes within the shocked region, and of the properties of the GRB jet and surrounding circumburst environment. A bright and well-monitored GRB such as GRB 130427A therefore provides an opportune dataset to test these simplified model assumptions. Previous examples of nearby GRBs with extremely good data coverage are GRB 030329 and GRB 080319B (also referred to as the ‘the naked-eye burst’[4]), both of which showed complex afterglow light curves containing unexpected breaks and re-brightenings which cannot be explained by a single, forward shock emission component [70,130–132]. On the other hand, GRB 130427A showed a comparatively smooth decay that could be well fitted by a single reverse and forward shock emission component [123,126,127,133,134]. Such comparatively simple afterglow behaviour enables the detailed properties of the forward shock to be scrutinized, such as the small (less than 0.15) fraction of relativistically accelerated electrons [124,126], which is typically assumed to be 1, the time dependence of the fraction of energy in the shocked electrons and in the magnetic field [124], which is nominally treated as static, and the subsequent movement of the synchrotron cooling frequency, which was slower than predicted by basic theory [123,124], but in agreement with previous indications that the cooling frequency remains at high energies (greater than 0.003 keV) for much longer than expected [135]. Apart from the microphysical parameters that define the conditions within the shocked region, the GRB afterglow evolution depends on the total energy in the ejecta and the density and density profile of the circumburst medium. There is universal agreement in the literature that GRB 130427A had an unusually low circumburst medium density ($n < 10^{-3}$ particles cm$^{-3}$), and this very likely had a wind-like density profile with a radial dependence approximately $r^{-2}$ (but see [125,126]). The low density medium probably gave rise to the long-lived reverse shock [127]. Although GRB 130427A was exceptional in its luminosity (especially compared to other long GRBs at $z < 0.5$), the derived properties of the shocked region and circumburst environment could have important implications for other long GRBs where such detailed analysis is not possible.

The afterglow evolution of most GRBs with multi-band data is generally found to be more consistent with a forward shock moving through a constant density profile medium [136], which goes against expectations for a massive progenitor star. However, these findings are mostly based on only optical and X-ray afterglow data. When very broadband, simultaneous data are available, extending as far as to the radio, as in the case of GRB 130427A, then a wind-like density profile is often preferred as in the case of GRB 130427A, and other notable GRBs such as GRB 080319B [70] and GRB 121024A [137]. These findings imply that in order to accurately disentangle the various microphysical, dynamical and geometrical factors contributing to the observed afterglow evolution, it is imperative to have full spectral coverage of the synchrotron afterglow spectrum during the afterglow’s evolution. Future exceptional

[4]GRB 080319B was a very luminous GRB at $z = 0.937$. It had the brightest optical emission ever detected, with a peak visual magnitude $V = 5.4$ mag, which could have been detected by the naked eye, thus giving this GRB its name.

GRBs such as GRB 130427A will provide further understanding on the GRB–SN connection and the origin of the large range in GRB luminosities.

### 2.4.2. Ultra-long gamma-ray bursts and their SNe: the case of GRB 111209A/SN 2011kl

The recently proposed class of ultra-long GRBs (§2.1) provides yet further opportunities to explore the final end stages of massive stars, and the diverse observational signatures that they give rise to. The very long duration prompt emission light curves of ultra-long GRBs (lasting for tens of thousands of seconds) implies that the central engine is active for approximately 100 times longer than in typical long GRBs and their X-ray light curves are also markedly different from standard long GRBs [29,34,138–141]. One of the first emission mechanisms suggested was the TDE of a main sequence star by the galaxy central supermassive black hole [38]. However, the $\gamma$-ray emission of such events last for an order of magnitude longer than in the case of ultra-long GRBs, and TDEs are an order of magnitude less luminous [29]. To account for this, the tidal disruption of a white dwarf by a low-mass central black hole (i.e. $M_{BH} < 10^5\ M_{\odot}$) has also been suggested [29,38]. The host galaxies of three GRBs classified as ultra-long by Levan *et al.* [29] are low luminosity, compact galaxies, and the location of the transients are consistent with having arisen at the galaxy nucleus, where the central black hole would reside. However, GRB 130925A, which had high energy emission lasting for approximately 20 ks and a luminous X-ray light curve and strong X-ray flaring [141] reminiscent of the three ultra-long GRBs proposed by Levan *et al.* [29], is spatially offset from the host galaxy nucleus, effectively ruling out a TDE origin. The core collapse of a low-metallicity blue supergiant into a black hole (BSG) has also been proposed as a possible progenitor channel [29,34]), although the near-solar metallicity of the host galaxy of GRB 130925A would also disfavour such a model [142]. As is the case with the collapsar model of normal long-duration GRBs, a BSG progenitor requires a low-metallicity star to maintain sufficient angular momentum at the time of core collapse to form the internal engine that powers the GRB [34,35]. This is due to the decrease in mass loss through line-driven winds at lower stellar metallicities.

Although these host galaxy observations provide some insight on the origin of ultra-long GRBs, the inferred progenitor properties are nevertheless indirect. As was the case with normal long-duration GRBs, a fairly recent and exciting revelation came with the spectroscopic detection of an SN coincident with the ultra-long GRB 111209A [39]. Other ultra-long GRBs have either been too far (e.g. GRB 121027A at $z = 1.774$ [29]) or in the case of GRB 130925A, at $z = 0.347$, a large amount of host galaxy dust that produced a visual extinction of $A_V = 5.3$ mag [143], almost fully extinguished the optical afterglow, and would certainly have blocked any emission from an underlying SN. Two intermediate ultra-long GRBs, GRB 101225A at $z = 0.847$ [27,28] and GRB 111209A at $z = 0.677$ [29,39], showed evidence of flattening in their optical and NIR light curves at 10–20 days after the GRB, indicative of additional emission from a rising SN. However, this was only spectroscopically confirmed in the case of GRB 111209A with an *X-shooter* observation taken around 20 days after the GRB, close to the peak of the associated SN 2011kl. Similarly to GRB 030329/SN 2003dh, the contribution from the GRB and host galaxy emission had to be removed in order to extract the spectrum of SN 2011kl, and in fact initially the SN was not found, highlighting the complexity of such analysis [29].

The detection of an SN coincident with the ultra-long GRB 111209A [39] rules out a TDE origin, and the lack of hydrogen features in the spectrum of SN 2011kl also disfavours the BSG interpretation, which should give rise to a hydrogen-rich SN. Possibly most informative of all was the peculiar spectral shape of SN 2011kl, which was very blue and featureless, unlike other GRB-SNe observed (figure 1). Both its light curve and spectral properties are instead more reminiscent of the newly discovered class of superluminous supernova [144,145] than of Type Ic-BL SN associated with long GRBs. Further similarities to SLSNe was the good fit provided by a magnetar-powered central engine to the GRB light curve [39,103] instead of the more commonly assumed black hole-accretion disc central engine. However, at a peak absolute bolometric magnitude of −20.0 mag, SN 2011kl is an order of magnitude too dim to be considered an SLSN, and instead represents an intermediate class of SNe that may bridge the gap between SLSNe and the more common class of standard core-collapse SNe [146]. The universality of the connection between ultra-long GRBs and if not superluminous, then very luminous SNe of course needs to be substantiated with a larger sample of 'nearby' ultra-long GRBs for which spectroscopic follow-up is feasible. Nevertheless, the detection of the unusual SN 2011kl coincident with GRB 111209A has provided the first concrete evidence of a common core-collapse origin for long and ultra-long GRBs alike, and this event illustrates how an SN detection coincident with a GRB greatly enhances our ability to discriminate between progenitor models.

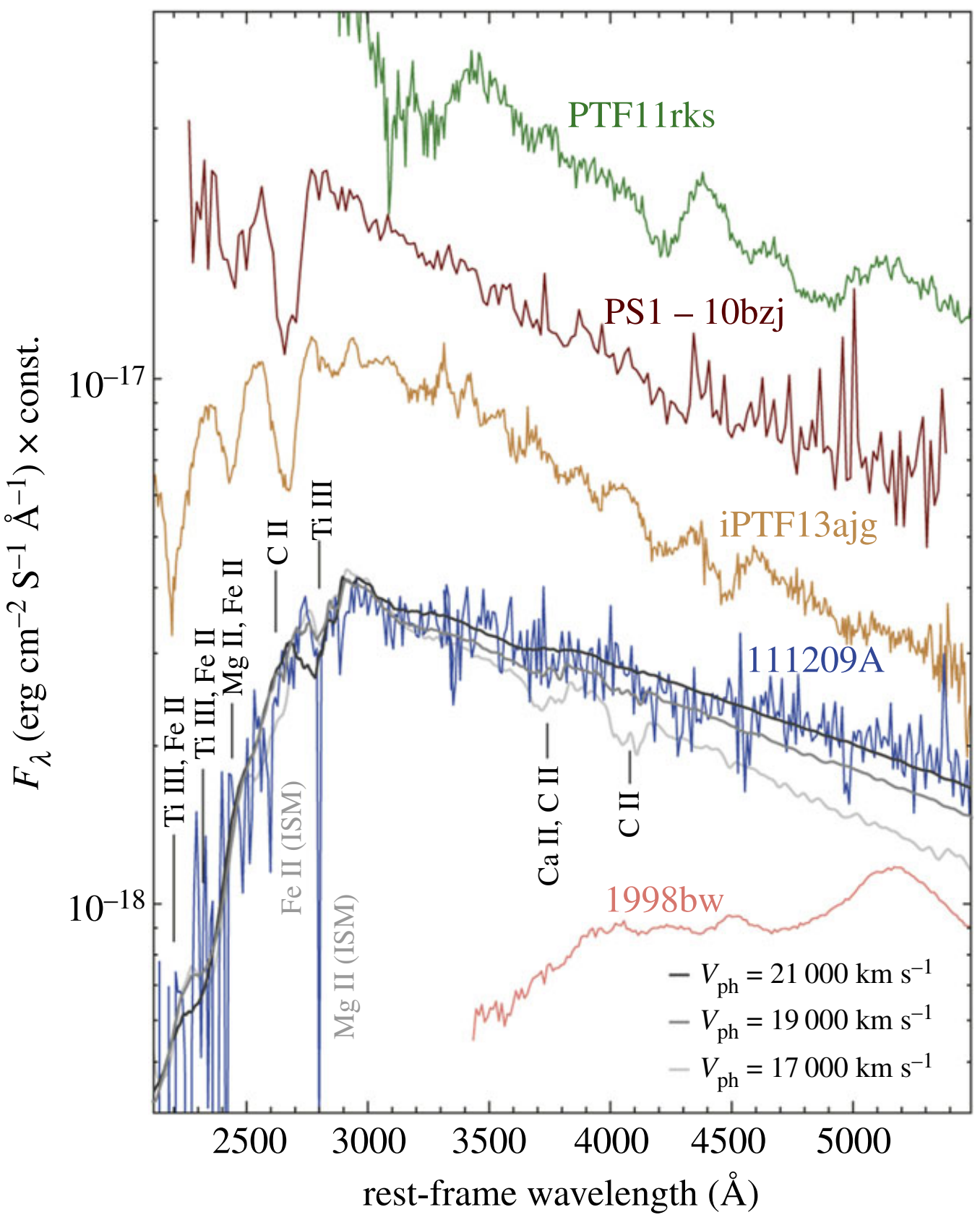

**Figure 1.** The *X-shooter* spectrum of SN 2011kl, associated with GRB 111209A, compared to the archetypal GRB–SN, SN 1998bw (pink), and spectra of three well-known SLSNe (top three curves). The three solid lines correspond to synthetic spectra with photospheric velocities of 21 000, 19 000 and 17 000 km s$^{-1}$, going from darkest to lightest grey. The flux scale is correct for SN 2011kl and SN 1998bw, but all other spectra have been arbitrarily shifted for clarity [39].

## 2.5. Short gamma-ray bursts and kilonovae

Whereas the *Swift* mission has accelerated our understanding of long-duration GRBs, the progress on short GRBs has been substantially slower. The rapid and precise localizations of short GRBs enabled with *Swift* has resulted in an increase in the number of optical, and in particular X-ray afterglow detections, and the first optical afterglow spectrum of a short GRB was finally acquired for GRB 130603B, providing a firm redshift, and subsequent secure host galaxy identification [147,148]. Nevertheless, short GRBs have far less luminous afterglows than long GRBs, and subsequently, the detection rate of short *Swift* GRB optical afterglows is only approximately 30% compared with approximately 50% for long GRBs.[5] The continual difficulty in obtaining optical afterglow spectra of short GRBs has limited progress in this field, with redshifts relying on sometimes uncertain host galaxy associations, and absorption spectroscopy from the intervening circumburst and ISM being largely unavailable. As such, information on the progenitors of short GRBs remains mostly indirect. For example, the lack of any core-collapse SNe associated with short GRBs [26,149–151], the typically large offsets between the GRB position and galaxy nucleus [152–155], and the frequent association with galaxies that have no ongoing star formation ([24,25,156] and references therein) have all been taken as evidence in support of a compact binary merger progenitor scenario.

One of the most promising prospects of confirming the binary merger progenitor model is through the detection of the predicted ‘kilonova’ that is expected to occur simultaneous to a short GRB. It has long been hypothesized that the binary merger of an NS–black hole, or NS–NS system will produce significant quantities of neutron-rich radioactive species that decay to form transient emission with peak luminosities up to 1000 times brighter than a nova, hence the name ‘kilonova’ [157–161]. Such emission should also accompany any short GRB if the progenitor model is correct. However, at the typical redshift of short GRBs of $z \approx 0.5$, any prospect of seeing the comparably dim kilonova requires deep follow-up observations. Indeed, early attempts to detect a signature of kilonova emission in the optical afterglow light curves of short GRBs were unsuccessful [162].

[5]Statistics calculated using table from http://www.mpe.mpg.de/~jcg/grbgen.html.

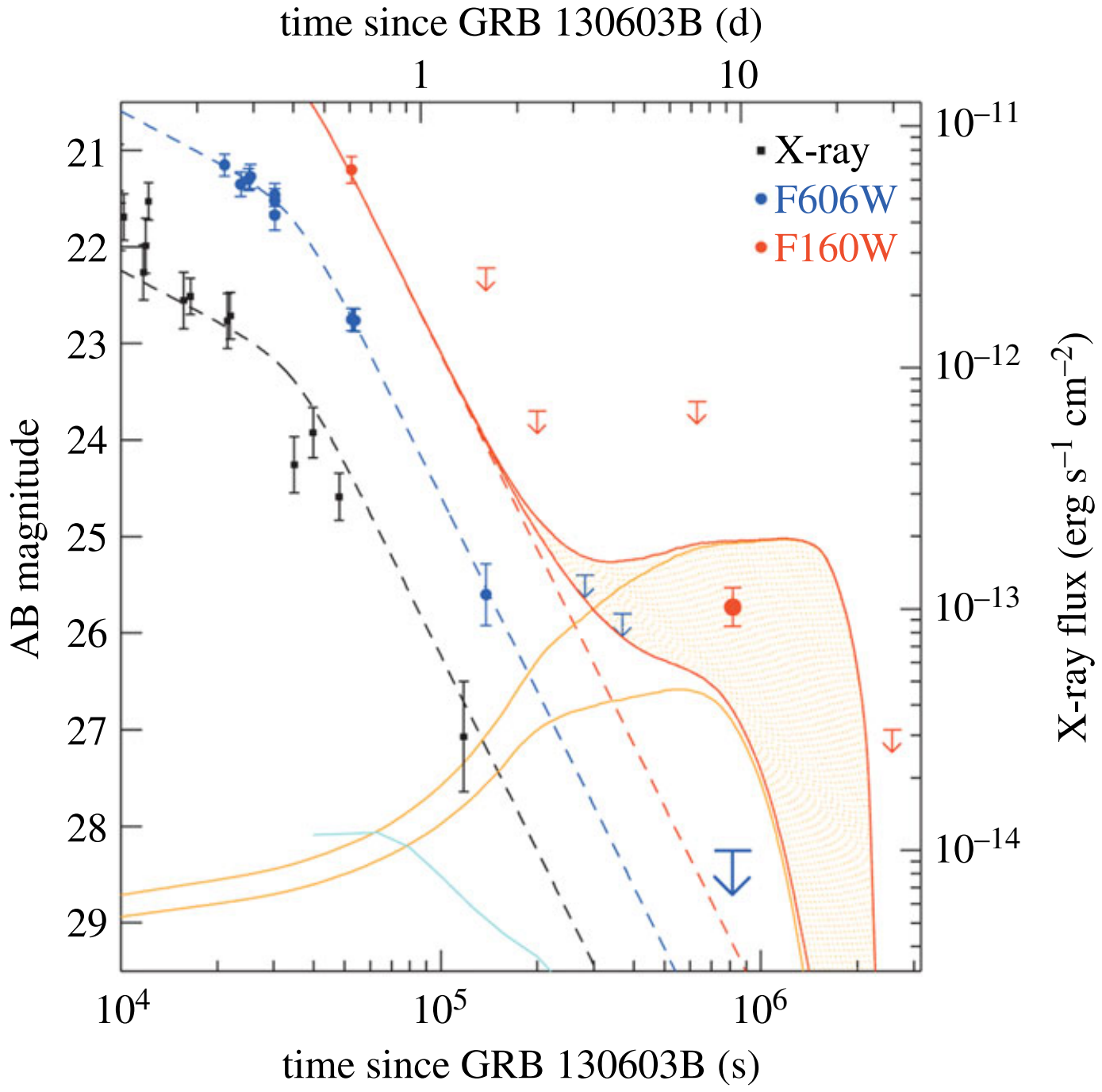

**Figure 2.** NIR, optical and X-ray light curve of the short GRB 130603B [166]. The corresponding optical AB magnitude and X-ray flux are given on the left and right axes, respectively. The optical (blue) and X-ray (black) afterglow light curves are well fitted with the same broken power law. Assuming the same post-break afterglow decay rate in the NIR (red), it is clear that an additional component is necessary to account for the detection at approximately 9 days after the GRB. The lower and upper orange curves correspond to predicted kilonova NIR light curves calculated from models with respective ejected masses of $10^{-2}$ $M_{\odot}$ and $10^{-1}$ $M_{\odot}$ [164], and the solid red curves are the corresponding emission from the GRB afterglow plus the kilonova. The cyan curve is the brightest optical emission predicted from an r-process kilonova, which is orders of magnitude dimmer than the GRB afterglow.

The first indication of a kilonova following a short GRB finally came from the extensive follow-up of GRB 130603B, which was one of the nearest and brightest short GRBs ever detected, and also the first short GRB with an optical afterglow spectrum. Of immense relevance in this follow-up campaign were the recent theoretical calculations at the time that showed the opacities of the heavy r-process elements created during the merger to be orders of magnitude larger than in iron-rich supernova ejecta, thus producing much dimmer and redder kilonovae than previously believed [163–165]. In the light of these developments, deep observations of GRB 130603B were carried out in the NIR bands using the Hubble Space Telescope (HST), and a re-brightening was detected approximately 9 days after the GRB (figure 2; [166]). This is the most direct evidence yet that short GRBs arise from compact-object mergers, and presents the exciting prospect of deriving some physical properties related to the binary system, such as the ejecta mass. Two further claims have since been made of a kilonova (or micronova) detection in the re-analysis of the NIR light curves of two older short GRBs (GRB 050709 [167] and GRB 060614 [168]). Nevertheless, as was initially the case with the first SN detection associated to a long GRB, how ubiquitous kilonovae are among short GRBs remains to be seen. A recent promising candidate of a short GRB with a detectable kilonova was GRB 160821B, which was spatially coincident with a galaxy at $z = 0.162$. However, in this case deep HST NIR observations revealed no re-brightening, which implies that any underlying kilonova must have been a factor of five dimmer than the kilonova associated with GRB 130603B [169].

If short GRBs are produced from the merger of two compact objects, then they should also be a significant source of gravitational waves. However, despite the recent detections of gravitational waves from black-hole/black-hole mergers [170], the prospect of detecting gravitational waves from short GRBs remains very small. Short GRBs are very rare, and none have been detected within the few hundred mega parsecs out to which gravitational wave facilities are sensitive to compact binary mergers [171,172]. A more realistic expectation is that a kilonova is detected coincident with a gravitational wave event from a binary merger system that is unassociated with a GRB. Only a small fraction of compact binary mergers are likely to produce short GRBs, and these will only be detected if the beamed emission is pointing towards us. Kilonovae, on the other hand, are expected to be ubiquitous with compact binary mergers, and their emission is emitted isotropically, greatly increasing the potential detection rate. Aside from the kilonova detected alongside GRB 130603B, our knowledge of kilonovae is purely theoretical. Acquiring a sample of kilonova multi-band light curves with known progenitor binary masses will help determine what range of kilonova properties may be expected from short GRBs and what physical properties should be inferred from future GRB-kilonova detections or non-detections.

# 3. The environments traced by long-duration gamma-ray bursts

From very early on in the field of GRB research, it was recognized that the immense luminosities released by long GRBs, and their association with massive stars, offered a powerful probe of distant star formation. Although attempts to use GRBs as standardizable candles have been thus far controversial and of limited use, their very bright prompt and afterglow emission provides an opportunity to detect regions of the Universe that would otherwise largely remain unseen. Notably, they single out regions of star formation independent of host galaxy luminosity, and their bright $\gamma$-ray emission can be detected out to the epoch of reionization, with the two most distant GRBs currently known having a redshift in the range $z = 8$–10 [173–176].

In view of this, there have been several investigations that have used the rate of long GRBs to study the cosmic star-formation rate (SFR) density at $z > 4$ [177–180]. Although GRBs are rare events, they can nevertheless provide novel insight on the contribution to the cosmic SFR from galaxies that are below the sensitivity detection limit of conventional surveys. The long GRB rate peaks at $z \sim 2$, which is similar to the peak of the cosmic SFR activity. Using a number of different techniques to normalize the observed long GRB rate to a corresponding cosmic SFR density at $z < 4$, the GRB redshift distribution has consistently implied a larger SFR density at $z > 4$ than is inferred from galaxy surveys. The still unclear details on the progenitor properties that give rise to a GRB mean that there is a large associated error on any GRB-derived cosmic SFR density. Conversely, mapping the cosmic SFR history from galaxy observations also has sources of error related to uncertain dust corrections, and more notably, to the validity of extrapolations of the galaxy luminosity function at high redshift [179,180].

There is now ample evidence indicating that long GRBs preferentially form in subsolar metallicity environments, although the functional form of this metallicity dependence is unclear. GRB progenitor models need to maintain sufficient angular moment at the moment of core collapse to form a black hole-accretion disc system, or alternatively a highly spinning magnetar, and this condition is more easily met at metallicities $\lesssim 0.3\ Z_\odot$ [181–183]. At higher metallicities, line-driven stellar winds remove significant angular momentum. However, the hard metallicity cut-off implied by progenitor models is not confirmed by observations. Another effectively unknown but important environmental factor is the stellar initial mass function (IMF) at high redshift. The progenitors of long GRBs are undoubtedly massive stars, and thus a skewed IMF to larger masses, as has been proposed to be in place in the younger Universe [184,185], would increase the rate of long GRBs per unit stellar mass [186]. In this section, I provide an overview of our current understanding of the environments traced by GRBs, and of the selection effects to be aware of in a GRB-selected galaxy sample. For a more detailed and very comprehensive review on the use of GRBs to study the cosmic SFR density, I redirect the reader to [187].

## 3.1. Heavily dust-extinguished gamma-ray bursts and their host galaxies

The first samples of long GRB host galaxies detected in the late 1990s and early 2000s were almost exclusively metal-poor, UV-bright, irregular dwarf galaxies [112,188], in line with the predictions of the progenitor models. Nevertheless, the typically long delay between the detection of the GRB and the start of optical follow-up observations, and more importantly, the general lack of infrared (IR) follow-up, introduced large selection effects against high-redshift events, and more notably, against heavily dust-extinguished GRB afterglows. With the launch of *Swift* and the simultaneous commissioning of several IR cameras on semi-robotic telescopes (e.g. PAIRITEL [189], GROND [190]), the extent of these selection effects became apparent, and it is now estimated that approximately 25% of long GRB afterglows are heavily extinguished by dust within the host galaxy, giving rise to a visual extinction (V-band) $A_V > 1$ mag [135,191]. This dust is predominantly located within the ISM of the galaxy. It was thus perhaps unsurprising to find that the host galaxy population for this sample of dust-enshrouded long GRBs are significantly more massive and chemically enriched than previous GRB host galaxy samples [142,192,193].

Whereas the host galaxies of optically bright GRBs have typical stellar masses of $10^9$–$10^{10}$ $M_\odot$ [114, 194] and metallicities that are well below solar [113,194], the host galaxies of dust-enshrouded GRBs have stellar masses more frequently in the range $10^{10}$–$10^{11}$ $M_\odot$, and there is now a notable fraction of GRB host galaxies with near-solar metallicities [142,194–197]. This greater diversity in the properties of GRB host galaxies implies that they are less biased tracers of star formation than previously believed, and the fraction of high metallicity cases all but rules out the notion of a hard metallicity cap [114,194]. This has led to the suggestion that either the environmental metallicity is not a dominant factor in the formation of long GRBs [198,199], or that there are at least two formation channels, one of which is more loosely dependent on progenitor metallicity [200]. It is, however, important to note that even with the inclusion of metal-rich host galaxies, the overall stellar mass and metallicity distribution of GRB hosts continues to be skewed to lower values relative to the general star-forming galaxy population out to $z \lesssim 3$ [114,200].

Despite efforts to develop progenitor models that are only weakly dependent on metallicity, theoreticians have found it hard to maintain high angular momentum right up until the moment of core collapse without placing some limit on the progenitor metallicity. A promising solution was to implement a binary progenitor system where the GRB progenitor is spun up by the companion star once they become tidally locked [201]. However, unless the stellar core and outer envelope are disconnected such that they evolve independently [202–204], most progenitor models still find that angular momentum is eventually removed from the core through stellar winds [205], and that some metallicity cap is thus necessary. In order to judge how severe the current contention is between observed and predicted metallicities, it is necessary to have a better observational constraint on the long GRB metallicity distribution. Whereas the metallicity distribution of long GRB host galaxies *relative* to the general star-forming galaxy population is broadly accepted to be skewed to lower metallicities, the absolute metallicity cap above which the long GRB production efficiency drops continues to be debated [113,114,206], ranging from approximately 0.4 $Z_\odot$ [206] to near-solar [114]. This discrepancy probably originates from differences in GRB host galaxy samples and datasets, and variations in the metallicity diagnostics applied. Some of the considerations that need to be kept in mind when inferring progenitor properties from the measured galaxy metallicity are discussed in the following section.

## 3.2. The metallicity distribution of long gamma-ray burst host galaxies

The large majority of host galaxy metallicities are measured from single-slit spectroscopy of the galaxy and applying strong emission-line metallicity diagnostics. A concern with such measurements is that they are averaged across the entire galaxy, and thus the metallicities measured may not be representative of the progenitor star metal abundance. Based on spatially resolved spectroscopic observations of M31, Niino *et al*. [207] found that spatial resolution better than 500 pc is required to measure the representative environmental metallicity for any transient event, and metallicities measured on spatial scales $\gtrsim 1.0$ kpc are generally more representative of the galaxy-averaged metallicity, irrespective of the metallicity within the transient natal region. Spatially resolved studies of the ionized gas within star-forming galaxies show variations in metallicity of approximately 0.3 dex about the mean [208,209], which could go a long way in resolving the apparent discrepancies between theoretical expectations and observations. Most of this variation in metallicity comes from galactic metallicity gradients. Thus, if the effect of poor spatial resolution is the principle cause for the high metallicities measured in some GRB host galaxies, we would then expect GRBs with high-metallicity host galaxies to reside predominantly in the outskirts of the galaxy. Using stellar mass as a proxy for metallicity, there is no compelling evidence that long GRBs with more massive and thus more metal-rich host galaxies reside at larger radii from the galaxy nucleus [210,211], although this is hard to verify for GRBs at higher redshift.

Another potential cause for concern is that of chance alignment with an unrelated galaxy along the line of sight to a GRB. This problem arises when a redshift from the GRB afterglow has not been attained, and thus the association between the GRB and a nearby galaxy when projected on the sky cannot be confirmed. Although most host galaxies confidently identified in this way have a very small chance probability of being unrelated to the GRB (i.e. GRB error circle is small and often the nearby galaxy is also relatively bright), there are now sufficient GRBs with spectroscopically unconfirmed galaxy associations that we would expect at least some of these to be incorrect. One such example was the previously considered canonical supersolar metallicity host galaxy of GRB 020819B that was only recently found to be a foreground galaxy at $z = 0.41$, with the GRB probably occurring at $z = 1.96$ [212]. In a similar vein, increasingly sensitive observations with the Very Large Array (VLA) and IR data taken with the *Herschel* space telescope are also ruling out previous claims that some GRB host galaxies are very highly star forming, ultra-luminous IR galaxies and submillimetre galaxies [213,214]. Nevertheless, although these new revelations may eliminate a few of the massive and metal-rich host galaxies previously claimed, the fraction of long GRB host galaxies with measured metallicities that are far higher than theoretical expectations remains notable.

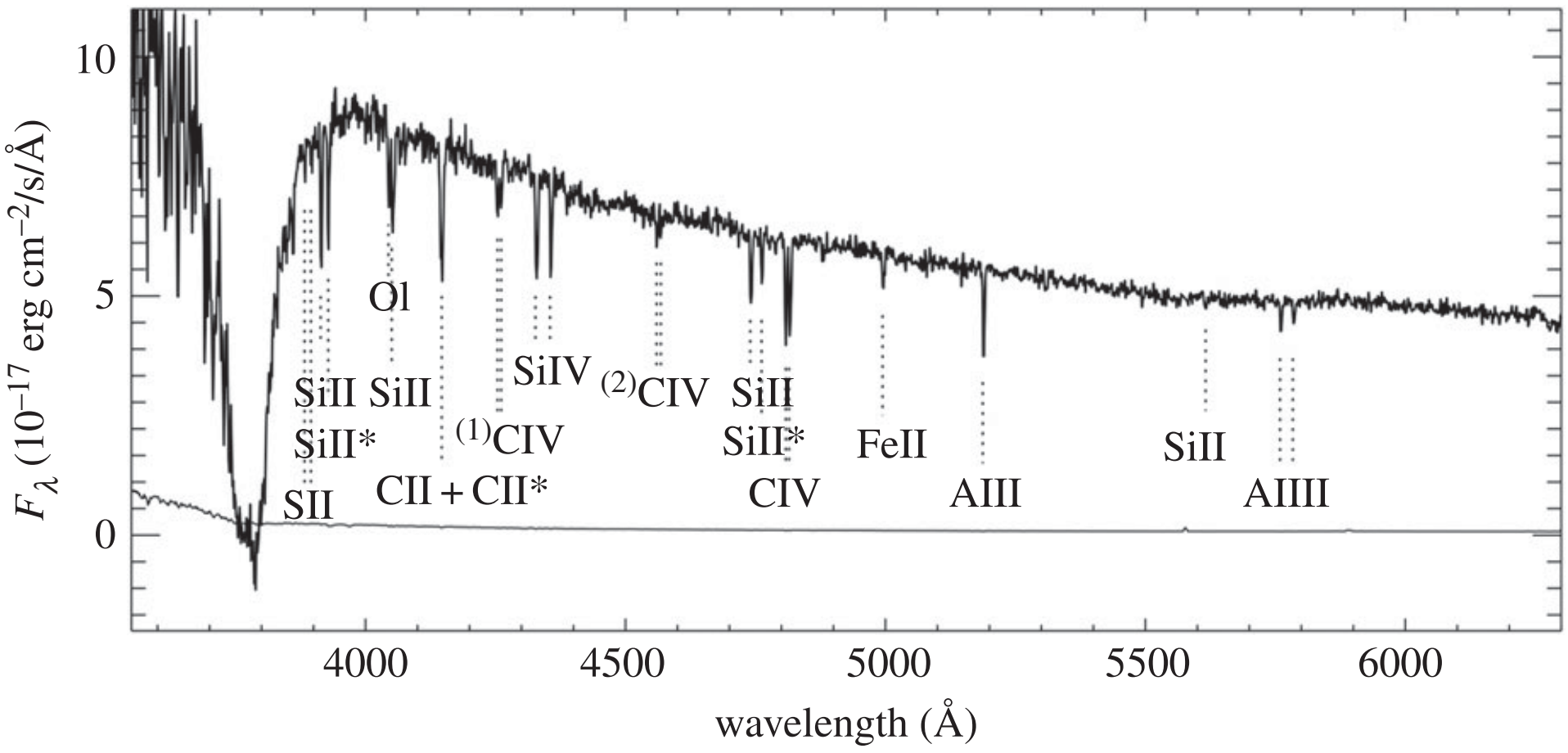

**Figure 3.** Observer frame VLT/FORS2 spectrum of the afterglow of GRB 090926A [221], clearly showing the Lyman-$\alpha$ absorption feature with $N_{\rm HI} = 10^{21.73\pm0.07}\,{\rm cm}^{-2}$, centred at 3800 Å, and numerous low and high ion metal absorption lines at a common redshift of $z = 2.1062 \pm 0.0004$. The best-fit absorption-derived metallicity using all available low-ion metal absorption lines is $Z = 10^{-2.37\pm0.16}\,Z_\odot$ [219]. Reproduced from Rau *et al.* [221] (Copyright 2010 AAS).

A more fundamental problem in GRB host galaxy metallicity measurements is likely to lie in the strong emission-line diagnostics used to derive the galaxy metallicity. These employ various metallicity-dependent relations between collisionally excited lines and recombination lines, predominantly oxygen, nitrogen and hydrogen. Depending on which emission lines are available, and on personal preference, the metallicity diagnostics may either use temperature-sensitive auroral emission lines, 'strong' emission lines (e.g. H$\alpha$, H$\beta$, [OII], [OIII], [N II]) that are calibrated against temperature-based diagnostics, or they are derived from theoretical photoionization models. The uncertainty in these various metallicity diagnostics is exemplified in the little agreement that there is between them, which for the same galaxy can vary by up to 0.7 dex [215]. Their validity at high redshift ($z > 1$), where most GRBs lie, is all the more uncertain, where the conditions of the ISM are known to differ considerably to conditions present in local galaxies.

A further, very compelling illustration of the uncertainty prevalent in emission-line metallicity diagnostics is in how they compare to the more direct gas-phase metallicities that can be measured from the hydrogen and metal absorption imprint left on the GRB optical afterglow. At high redshifts ($z \gtrsim 1.7$) absorption from neutral hydrogen within the GRB host galaxy is redshifted into the bandpass of UV/optical spectrographs, and typically reveal large column densities of neutral gas within the host galaxy corresponding to damped Lyman alpha (DLA) systems, formally defined as having $\log(N_{\rm HI})\,{\rm cm}^{-2} > 20.3$. Ionization simulations show that at these high H I column densities, hydrogen is largely shielded from UV radiation, rendering negligible ionization corrections [216,217]. Apart from the typically small correction for dust depletion of metals [218,219], and the negligible fraction of neutral hydrogen in molecular form [220] (see §4.4), the ratio of singly ionized metal lines to neutral hydrogen then gives a direct measure of the metallicity (figure 3) [219,222]. Even in the case where a notable fraction of metals is locked up in dust, or a significant amount of hydrogen is in molecular form, this can be measured rather than simply assumed. Although the overlap in GRB host galaxies with metallicities measured from both emission line and absorption lines is limited to a handful, the difference in the two methods is clear when looking at the host galaxy metallicity distributions derived from the two techniques, with absorption-based metallicities being systematically lower by up to an order of magnitude for certain emission-line diagnostics. Importantly, only a very small fraction of GRB host galaxies have absorption-based metallicities that lie above the theoretical 30% solar threshold imposed by most progenitor models (figure 4), which is in stark contrast to the distribution from emission-line-derived metallicities.

## 3.3. The very local environment of gamma-ray burst host galaxies

The compilation of 'complete' and unbiased GRB host galaxy samples, and their properties *relative* to other star-forming galaxy populations is possibly the most effective way of assessing the selection effects present in GRB-selected samples, irrespective of how well we understand the intricacies of the GRB progenitor [114]. Nevertheless, the number of detected GRB host galaxies samples remains relatively small when compared with most other galaxy samples,[6] and intrinsic scatter thus limits how well the distribution of host galaxy properties at a given redshift, say, can be determined.

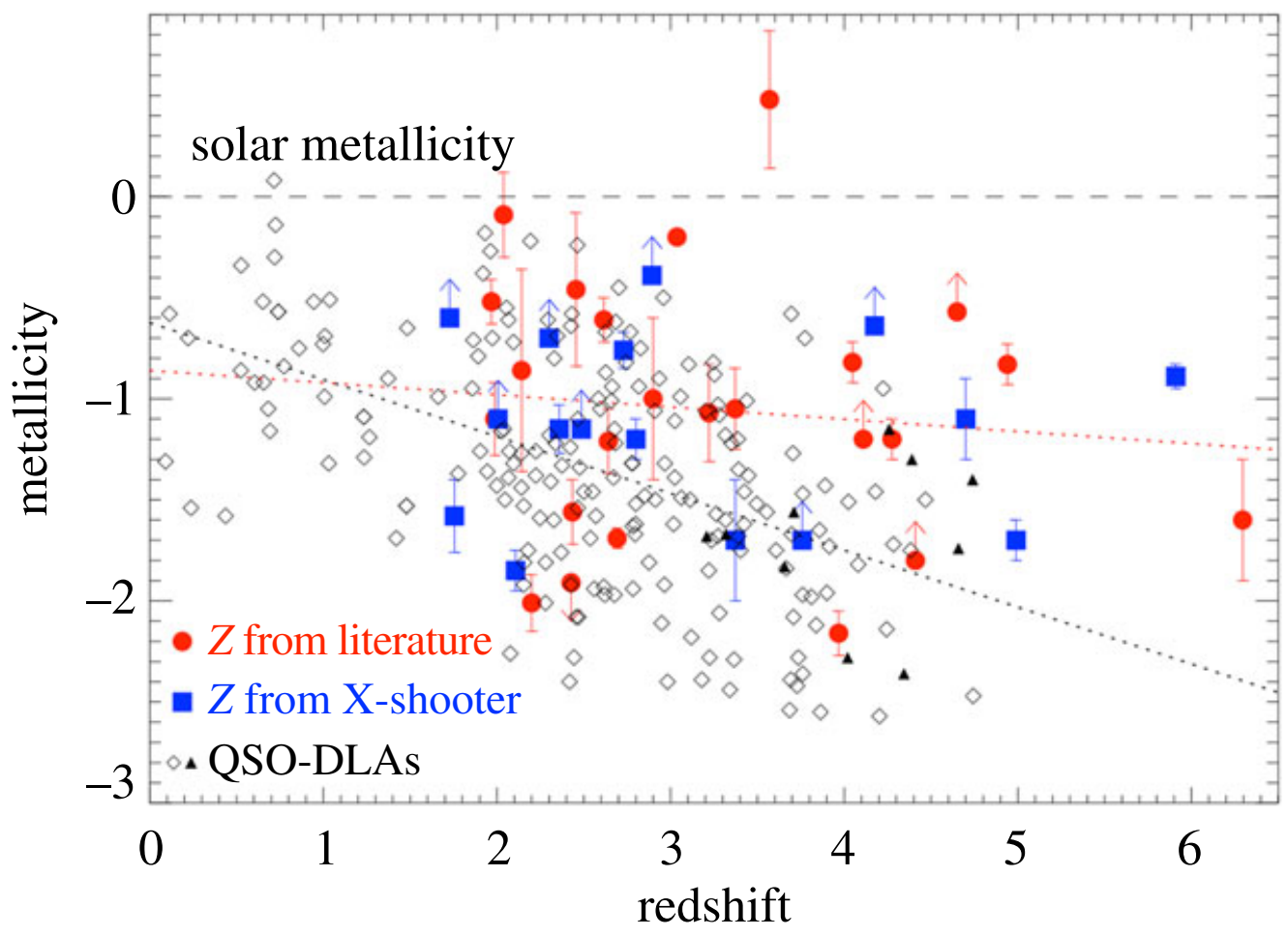

**Figure 4.** GRB afterglow and QSO–DLA absorption-derived metallicities versus redshift, adapted from [223]. The sample of GRB absorption metallicities measured using data from *X-shooter* (blue) and other spectrographs (red) predominantly lie below solar metallicities out to $z \sim 6$, and they are also mostly below the 0.3 $Z_\odot$ metallicity threshold typically imposed by GRB progenitor models. The QSO–DLAs have typically lower metallicities than GRB host galaxies, which is as expected given that QSO–DLA lines of sight typically cross the galaxy circumgalactic medium, whereas GRB lines of sight intersect the star-forming regions deep within their host galaxies.

Having a clear picture of the properties of long GRB progenitors and of the environments that they trace will undoubtably strengthen their use as probes to star formation. A very powerful way of studying GRB progenitors is through spatially resolved spectroscopic observations of their host galaxies using sensitive integral field unit (IFU) instruments. IFU observations have been possible for over a decade. However, the recent commissioning of the orders of magnitude more sensitive Multi-Unit Spectrograph Explorer (MUSE; [225]), mounted on the European Southern Observatory's Very Large Telescope (ESO/VLT), has revolutionized the spatial scales on which galaxies can be studied. MUSE has a 1 arcmin$^2$ field of view with a spatial resolution that is seeing limited,[7] and a spectral resolving power $R$ that ranges between 1500 and 3000 over the wavelength range 465–930 nm. Below I summarize some of the main results from high-quality MUSE data of the two nearest GRB host galaxies to date; that of GRB 980425 at $z = 0.0085$ (luminosity distance $D_L = 39.1$ Mpc) and GRB 111005A at $z = 0.0133$ ($D_L = 59.8$ Mpc).

MUSE observations of the comparatively nearby host galaxies of GRB 980425 and GRB 111005A provided an effective spatial resolution of 160 pc and 270 pc, respectively, enabling individual H II region-scales to be studied. This is unprecedented for a GRB host galaxy, and the sensitivity of the MUSE data provide extensive coverage of the hot gas within the host galaxies, even in the less star-forming regions. The host galaxy of GRB 980425 had been observed with the previous generation IFU instrument mounted on VLT/VIMOS (the Visible, Multi-Object Spectrograph; [226]), although the lower sensitivity provided coverage of only the brightest emission regions of the host galaxy, producing patchy data maps [227]. Even in these sampled host regions, the dust corrections were inaccurate, possibly as a result of imprecise flux calibration or stellar Balmer absorption corrections [209,227], which affect the accuracy of the derived SFR and host galaxy metallicity maps.

For the purpose of GRB progenitor models, two of the most relevant environmental properties are the stellar age and the metallicity within the *local* environment of the GRB, which are optimally studied with IFU observations. For example, it has been speculated whether GRBs with high-metallicity host galaxies may reside within metal-poor regions of their host, perhaps due to metallicity gradients across the galaxy. Such a hypothesis is challenging to verify, given the generally low spatial scales that are available for predominantly moderate mass, high-redshift GRB host galaxies. In the case of GRB 980425, the host galaxy had a subsolar metallicity, and the MUSE data show that the metallicity at the GRB position was lower than the galaxy average, in general, but also for the same projected distance from the galaxy centre. This is consistent with previous results based on VIMOS [227] or single-slit spectra [228], but the more accurate and complete MUSE coverage of the galaxy provides a more comprehensive view of the environmental conditions at the GRB position relative to the rest of the galaxy.

[6]The largest 'complete' GRB host galaxy sample is SHOALS [224], which contains 119 galaxies, compared to the tens of thousands common in galaxy surveys.

[7]MUSE was recently equipped with an adaptive optics facility that will be available to the community from April 2018, providing diffraction-limited observations with initially 0.2 arcsec pixels that will be later improved to 0.025 arcsec pixel scales.

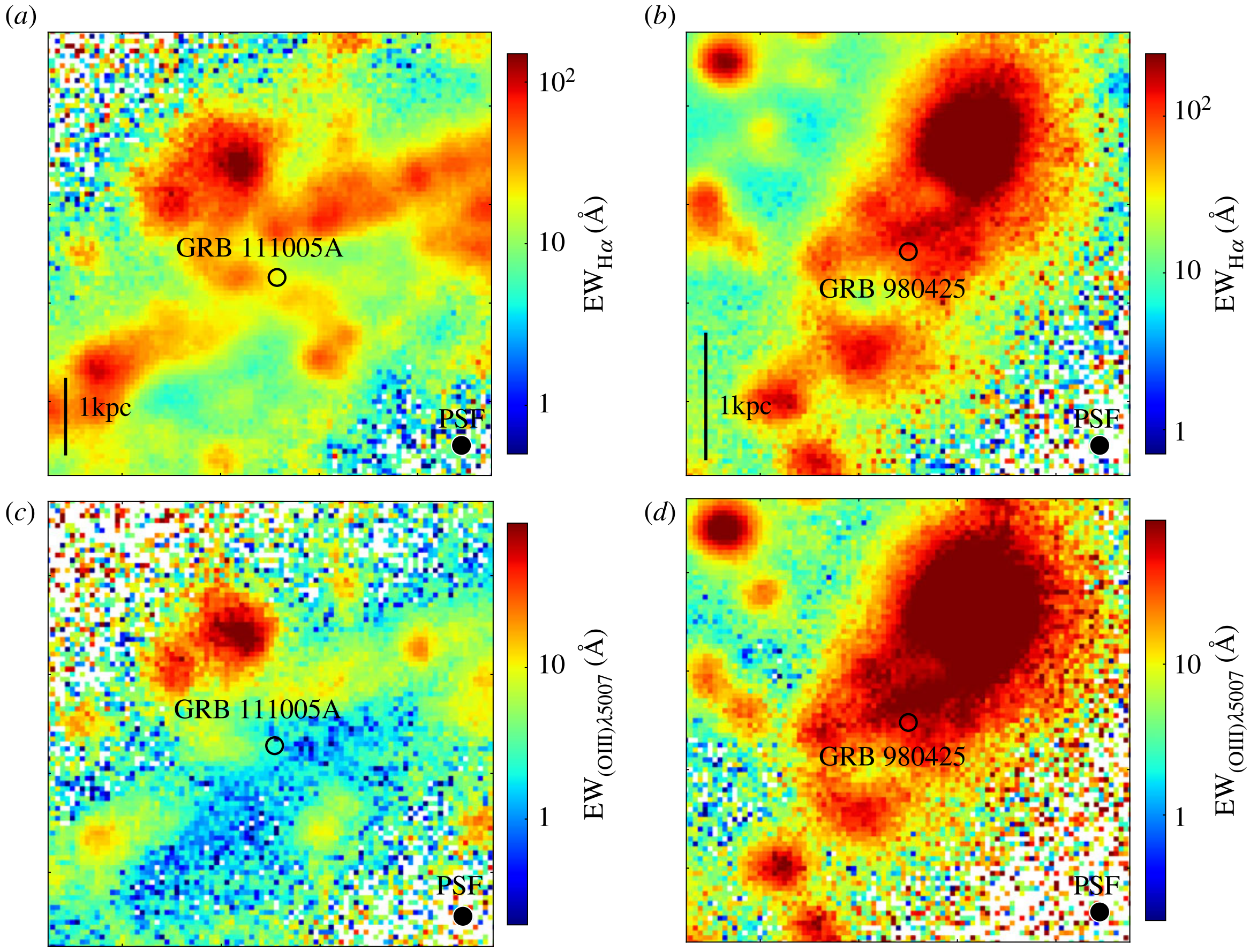

**Figure 5.** MUSE data maps of the equivalent width of H$\alpha$ (*a*) and [O iii] (*b*) in the nearby environments of GRB 111005A (*c*) and GRB 980425 (*d*). The position of the GRB is shown in each panel, which are approximately 20 × 20 arcsec in size. This corresponds to a physical scale of around 5.5 × 5.5 kpc for GRB 111005A, and 3.5 × 3.5 kpc at the redshift of GRB 980425. The spatial scale is indicated in (*a*,*b*). The effective spatial resolution is given by the point spread function (PSF), shown in the lower right corner of each panel, which in all cases has a full width half maximum of approximately 0.9 arcsec, or 160 and 270 pc at the redshift of GRB 980425 and GRB 111005A, respectively. In all images, north is up and east is to the left.

Possibly the greatest diagnostic power of the MUSE data is the chance to measure the age of the stellar populations within individual H II regions. Single stellar population models show a relation between the H$\alpha$ emission line equivalent width and the age of the stellar population, although it is only valid to apply such models to data where individual H II regions are resolved, as in the case of the host galaxies of GRB 980425 and GRB 111005A. A notable property of the nearby environment of GRB 980425 is a star-forming region approximately 5 arcsec (approx. 860 pc) northwest of the GRB explosion site that shows emission-line features from Wolf–Rayet (WR) stars, which are young, massive stars that are the probable progenitors of long GRBs [22,182]. Given the link between GRBs and massive stars, it has been proposed that the GRB progenitor originated within the WR region, but was kicked out, possibly by the SN explosion of a binary companion star [228]. From the H$\alpha$ equivalent widths measured at the position of the WR region and of the transient explosion, Krühler *et al.* [209] derived a stellar age of less than 3 Myr at the WR region, and an age of 3–5 Myr at the GRB position. The young age of the WR region makes the runaway progenitor star scenario highly unlikely, with an unusually high kick velocity having to be implemented in order for the progenitor to travel the distance to where it finally exploded. If instead the more natural assumption is taken that the GRB exploded within its natal region, the measured stellar age then translates to a progenitor zero-age stellar mass in the range 25–40 $M_\odot$ [209], which is consistent with GRB progenitor model predictions.

In contrast with GRB 980425, the host galaxy of the peculiar, supernovaless GRB 111005A (see §2.4) has a near-solar metallicity, and the MUSE data show no indication that the GRB occurred within an especially metal-poor region of its host (figure 5) (M Tanga 2017, personal communication). Furthermore, the H$\alpha$ equivalent width within the region of the GRB is very small (approx. 15 Å compared with approx.

90 Å in the case of GRB 980425), and there is near to no [O III] emission at the GRB position, suggesting the overall absence of massive stars. This is consistent with the deep stellar Balmer absorption lines seen in the MUSE spectra at the GRB position, which are well fit by synthetic stellar spectra made up of a predominantly old stellar population with an age of 1–3 Gyr, and a 10–20% contribution from a 10–20 Myr old stellar population [229]. The *in situ* stellar ages are far older than what is expected from a GRB collapsar scenario, and from what is observed in the vicinity of GRB 980425. The nearest H II region to GRB 111005A lies almost 300 pc away, but even this H II region shows far less ongoing star formation than in the nearby environment of GRB 980425. The peculiar nature of GRB 111005A, which had no detected SN down to deep limits, and an unusual radio afterglow [119], possibly make it unsurprising that the host galaxy and local environment of this GRB are markedly different from other GRBs. The limits placed by the MUSE data on the underlying stellar population that gave rise to GRB 111005A provide useful constraints for future models that try and explain this transient, as well as other supernovaless GRBs.

# 4. Probing the interstellar medium in high-redshift, star-forming galaxies

Apart from signalling regions of very distant star formation, the very bright and broadband afterglow provides a truly unique view of the ISM within the star-forming host galaxy. Sight lines to quasi-stellar objects (QSOs) also light up the intervening material, and their constant illumination allows them to be revisited at any time. Where GRBs exceed as powerful probes of the high-redshift ISM is threefold: (i) the intrinsically featureless GRB multi-wavelength afterglow provides a very clear view of the absorption imprint left by intervening gas, metals and dust; (ii) the GRB line of sight delves deep into the host galaxy, unlike lines of sight to QSOs, which preferentially cross the circumgalactic medium of intervening galaxies; (iii) the GRB afterglow fades rapidly, which, although it has its disadvantages, also permits the host galaxy to be studied at a later stage, free from the bright glare of the background source.

From the absorption imprint left by intervening material on the GRB spectrum, it is possible to study the properties of the host galaxy dust [135,230–233], the ionization state and kinematics of various intervening absorption systems [234–236], and the chemical composition [222,223] and molecular gas fraction of the host galaxy ISM [237]. Acquiring comparable information from galaxy emission observations requires numerous facilities covering a broad wavelength range, and significantly larger observing times than the single 1–2 h exposures typically used for optical afterglow spectroscopy. Even then the sensitivity attainable through absorption-line observations far outweighs emission data for high-redshift, low-mass, metal-poor galaxies, which are common amongst the host galaxies of long GRBs.

The use of GRBs to study the interstellar conditions of distant galaxies is optimized if observations are taken over a broad wavelength range, and very soon after the GRB trigger, when the afterglow is still luminous. Broadband coverage enables the intrinsic afterglow spectral slope to be well constrained, and it maximizes the coverage of absorption features originating from the various components of the circumburst and ISM, not to mention any intervening material external to the host galaxy. The very rapid acquisition of optical and X-ray afterglow observations available with UVOT and XRT on-board *Swift* have been instrumental in providing very rapid, arcsecond positions to the GRB community for further follow-up observations. Furthermore, the very high XRT detection rate (approx. 95% [79]) and observations with multi-wavelength imaging instruments, such as the GRB optical and near-IR detector (GROND; [190]) on the Max-Planck Institute 2.2 m telescope in La Silla, and the UV, optical and near-IR (NIR) spectrograph *X-shooter* [238] on the Very Large Telescope in Paranal, have had a tremendous impact on the investigation of the properties of the host galaxy intervening material. In the following subsections, I review the principle results of the last decade within the research of GRB afterglow absorption studies.

## 4.1. Host galaxy gas and metal absorption

The sample of long GRBs with spectroscopic data covering the Lyman-$\alpha$ absorption trough is now of the order of 80, and a large fraction of these (approx. 85%) correspond to DLAs with neutral hydrogen column densities typically an order of magnitude larger than in QSO–DLAs (figure 6). The distribution of host galaxy neutral hydrogen column densities along GRB lines of sight peaks at $\log[N_{\mathrm{H\,I}}\,\mathrm{cm}^{-2}] = 21.6$, and although selection effects may alter the distribution somewhat, the peak column density is fairly robust. Selection effects against dusty lines of sight may impact the high column density end of the distribution, assuming that more dusty and massive host galaxies have correspondingly larger column densities of atomic neutral gas. This is supported by observations of the host galaxy of the very heavily dust-extinguished GRB 080607 ($A_V \sim 3.2$ mag [237]), which has the largest column density of

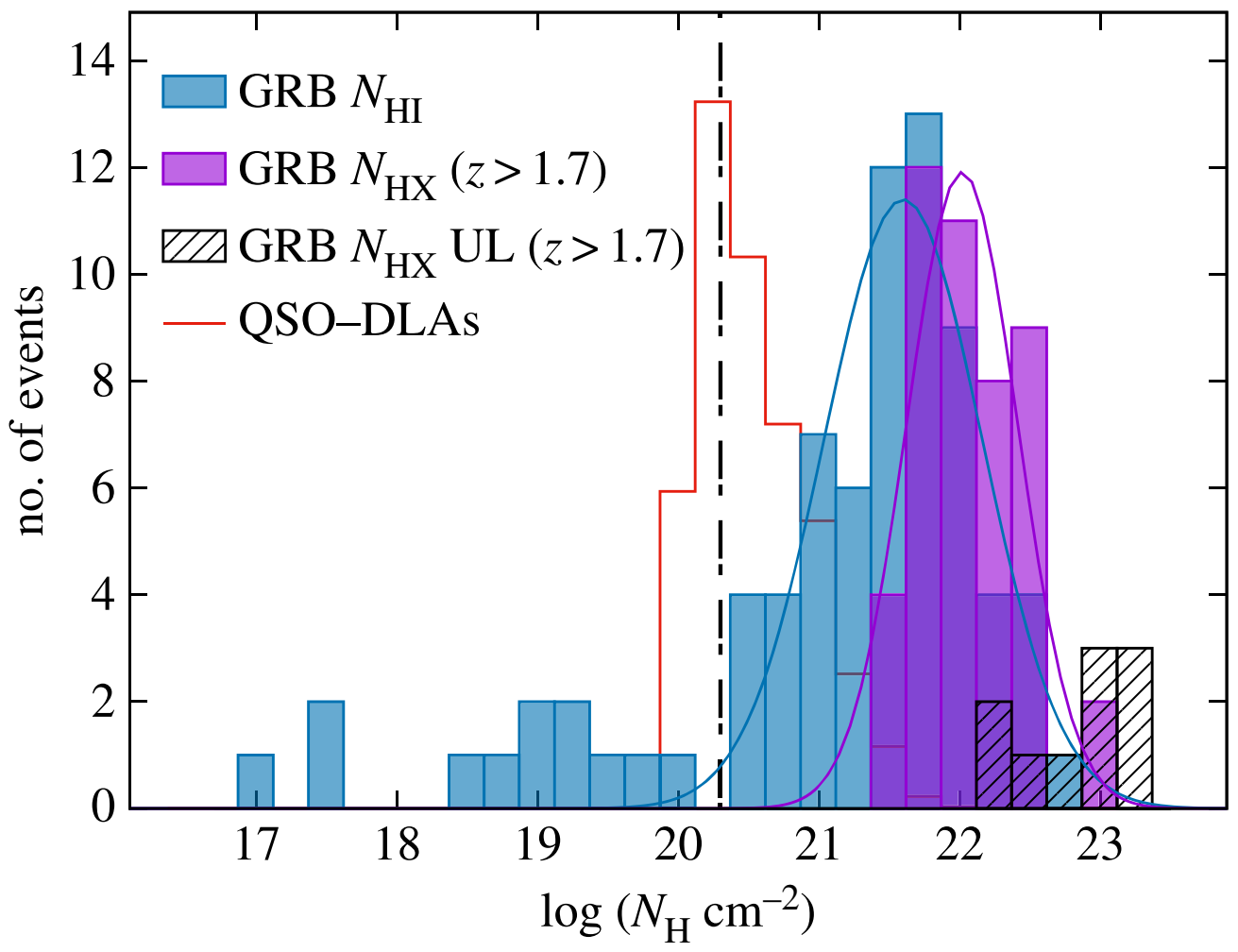

**Figure 6.** Distribution of host galaxy H I column densities measured from Lyman-$\alpha$ absorption along the line of sight to 77 long GRBs (blue), and equivalent H I column density (purple) and upper limits (black hashed) measured from the afterglow X-ray spectra when assuming solar metallicity for 56 long GRBs taken from [239]. The Lyman-$\alpha$-derived column densities are taken from [219,222,240–242]. The distribution of $N_{HI}$ and $N_{HX}$ column densities above $10^{20.3}$ cm$^{-3}$ (black dash-dot line) are well described by a Gaussian with peak $\log N_{HI}$ cm$^{-3}$ $= 21.6$ (blue line) and $\log N_{HX}$ cm$^{-3}$ $= 22.0$ (purple line), and respective standard deviations $\sigma_{N_{HI}} \sim 0.6$ and $\sigma_{N_{HX}} \sim 0.4$. As a comparison, the $N_{HI}$ column density distribution of 1426 QSO–DLAs and QSO-subDLAs from [243] is also shown (red), normalized to have the same number of events at the peak of the distribution as in the GRB $N_{HI}$ distribution.

H I gas ever measured in a GRB host galaxy ($\log N_{HI}$ cm$^{-3}$ $= 22.70 \pm 0.15$ [237]). Considering that up to approximately 25% of GRB lines of sight have $A_V > 1$ mag [193], the true $N_{HI}$ column density is likely to have an extended high-end tail. At the low end of the distribution it is possible that there is a selection effect against the publication of $N_{HI}$ upper limits or column densities with $\log N_{HI}$ cm$^{-3}$ $< 19.0$, where the absence of damped wings complicates the data analysis, and where it is also harder to derive a gas-phase metallicity. However, given that GRB afterglow spectra are frequently published for reasons other than the H I column density (e.g. time-varying lines [244,245], high-ion absorption features [234,246], high-quality afterglow spectra [219], investigations on 'complete' samples [240]), the effect of this bias is likely to be smaller than at the high end of the H I distribution.

The location of the neutral absorbing gas can be constrained by the survival of certain low-ion metal species, such as Mg I, which trace the neutral gas and place a lower limit of around 100 pc from the GRB [247]. More precise distances have been derived in a few cases where time-varying Fe II and Ni II fine-structure lines have been observed, and which are well modelled by the excitation of neutral gas located a few hundred parsecs from the GRB by the afterglow UV radiation [248]. In addition to the Lyman-$\alpha$ and low-ion metal absorption lines that trace the cold gas within the host galaxy ISM, highly ionized species such as O IV, C IV, Si IV and N V are also often detected in GRB optical afterglow spectra, which probe the hot gas ($T \sim 10^4$ K) within the ISM and circumgalactic halo [234], as well as possibly the GRB circumburst environment [246].

In contrast with the specific regions of gas that can be identified from UV spectra, the spectral resolution available from XRT X-ray afterglow observations limits the information that can be obtained on the location or ionization state of the absorbing material. It is common to detect soft X-ray absorption in excess of the Milky Way absorption [249,250], and for lack of further constraints, it is usual to ascribe this to photoelectric absorption from a solar metallicity, neutral gas cloud located at the GRB redshift. In figure 6, the host galaxy neutral hydrogen *equivalent* column density distribution, $N_{HX}$, from a *Swift* sample of 56 long GRBs at $z > 1.7$ is also shown (detections in purple and upper limits shown with a black hashed pattern). The peak of the $\log N_{HX}$ distribution is around 0.4 dex larger than the peak of the $\log N_{HI}$ distribution, and although the $N_{HX}$ upper limits push the distribution to slightly lower peak values, it is important to keep in mind that GRB host galaxies typically have a subsolar metallicity, which pushes the true equivalent H I column densities to larger values by a few tenths of dex.

This large discrepancy between UV/optical and X-ray absorption column densities was first realized in [251] using a sample of 17 *Swift* GRBs with both $N_{HI}$ and $N_{HX}$ measurements. It was later more robustly quantified in [236] by using singly ionized metal lines to measure the column density of neutral gas, which can then be directly compared to the X-ray absorption column density without requiring knowledge of the metallicity of the gas. The X-ray afterglow is predominantly absorbed by medium weight metals, such as oxygen and nitrogen, and the absorption cross section is a relatively weak function of the ionization state. One possibility is thus that the X-ray afterglow is absorbed by an additional and significant component of ionized gas that is transparent to the UV and optical afterglow [236,251]. High ions such as Si IV, C IV, N V and O VI, which are commonly detected in absorption in GRB afterglow spectra, can only account for approximately a further 10% of the X-ray absorbing gas [236]. The implication is thus that if the X-ray afterglow is predominantly attenuated within the host galaxy, the absorbing material has to be in an ultra-ionized state, presumably within a confined, dense region close to the GRB [236,252,253]. Alternative explanations to account for the X-ray absorption excess are that the X-ray afterglow is absorbed by large quantities of He I within the GRB natal H II region rather than by metals [254], or that the excess X-ray absorption stems from material external to the host galaxy, either within numerous discrete intervening systems [255,256], or within the warm and cold intergalactic medium, which must have a metallicity greater than 0.2 $Z_\odot$ [250,257].

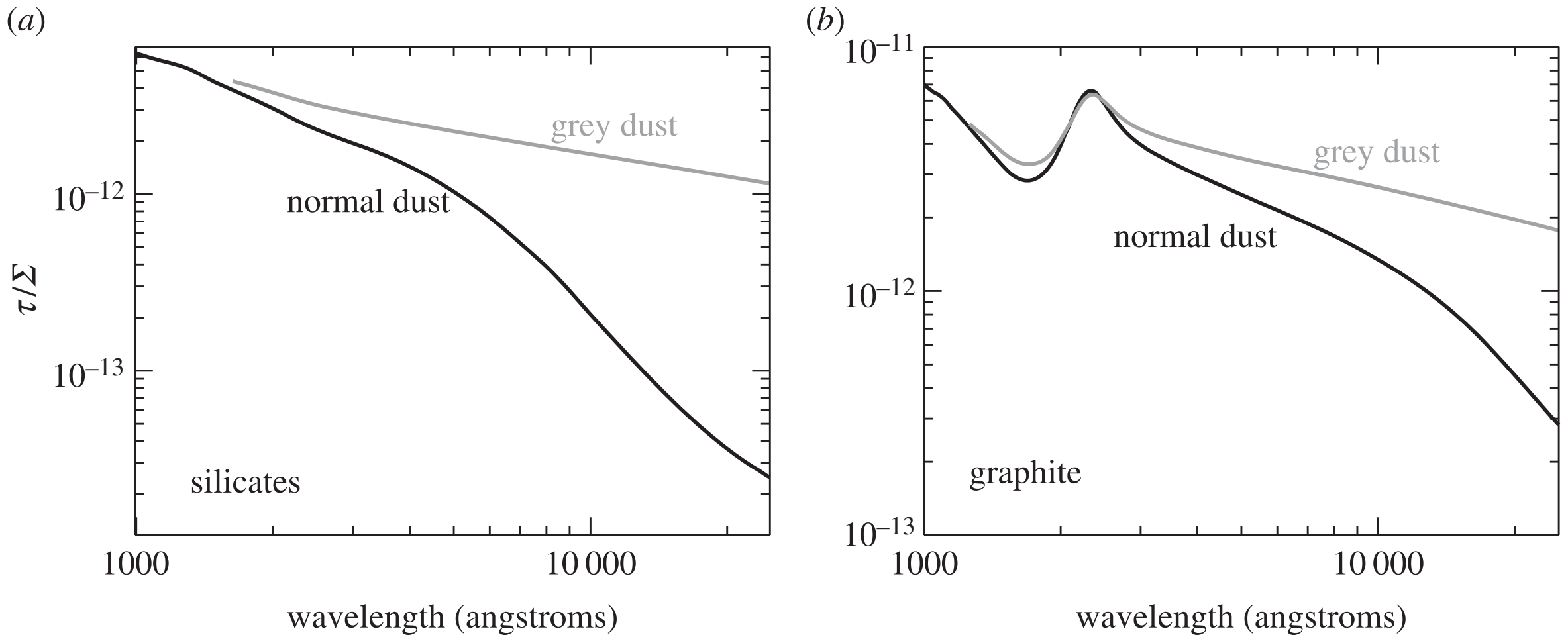

**Figure 7.** (*a*) Extinction curves for a population of silicate dust grains with a normal particle size distribution (black) with grain diameter in the range of $a = 0.005$–$0.25$ µm, and a grey distribution (or distribution skewed to large dust grains), with diameter in the range of $a = 0.005$–$10.0$ µm. The particle size distribution is of the form $(a/a_0)^{-3.5}$ [263]. (*b*) Same as left figure, but for a population of graphite dust grains. In both panels, the y-axis is the optical depth, $\tau$, divided by the column density of dust, $\Sigma$ (Credit: Adria Updike, NASA/GSFC).

UV spectra taken with the HST Cosmic Origins Spectrograph (COS) of nearby ($z < 0.5$) intervening O VI absorbers along QSO lines of sight show that the median metallicity of the warm intergalactic medium is 0.1 $Z_\odot$, which is too metal-poor to account for the X-ray excess [258], and the large number of intervening systems required to account for the order of magnitude absorption excess also seems unlikely [239]. Evidence to support an internal host galaxy origin include an observed positive trend between $A_V/N_{HX}$ and $A_V$ [192], very large columns of X-ray absorbing gas along the line of sight to nearby GRBs [239], and more recently, a correlation between $N_{HX}$ and host galaxy stellar mass, $M_\star$ [259]. This newly observed correlation between $N_{HX}$ and $M_\star$ implies that the GRB X-ray afterglow is predominantly absorbed by the host galaxy ISM, which is somewhat in contention with the results from [253], where it is shown that the hot gas component of the galaxy ISM is too diffuse to account for the X-ray absorption excess. In order to consolidate these two results, a large column density of X-ray absorbing material would have to not necessarily signify a large X-ray absorption excess. It may not be until the launch of the very sensitive X-ray mission *Athena* [260] that the origin of the X-ray absorption is conclusively resolved. Nevertheless, prior to this it should be possible to place more stringent constraints on the location of the X-ray absorbing gas with the use of detailed modelling of the different phases of the ISM, and of the impact of the GRB on the material along the line of sight, such as in [252,253].

## 4.2. Long gamma-ray burst host galaxy dust extinction curves

The single GRB sight line through the ISM of the host galaxy offers one of the only ways to study the dust extinction properties of distant star-forming galaxies. The dust *attenuation* properties of extragalactic star-forming galaxies has been studied by fitting the galaxy emission spectra with models containing varying amounts of dust and different dust distributions [261], but this is distinct from dust *extinction*. The dust attenuation of light by dust is subject to complex radiative transfer effects that reprocess the stellar light through numerous episodes of gas and dust absorption, emission and scattering, and is highly dependent on the geometric distribution of the dust, gas and stars [262]. Instead, the amount of dust extinction as a function of wavelength, referred to as the dust extinction curve, is only dependent on the dust composition and the dust size distribution, and studying these two properties provides some understanding on the dominant sources of dust production. This is illustrated in figure 7, where the left and right panels show the extinction curves produced from a population of purely silicate and graphite grains, respectively, with a 'normal' power-law grain size distribution (black), and a grain distribution skewed to large dust grains (grey). This latter distribution is also referred to as a 'grey' distribution due to the relatively weak dependence that the extinction law has on wavelength.

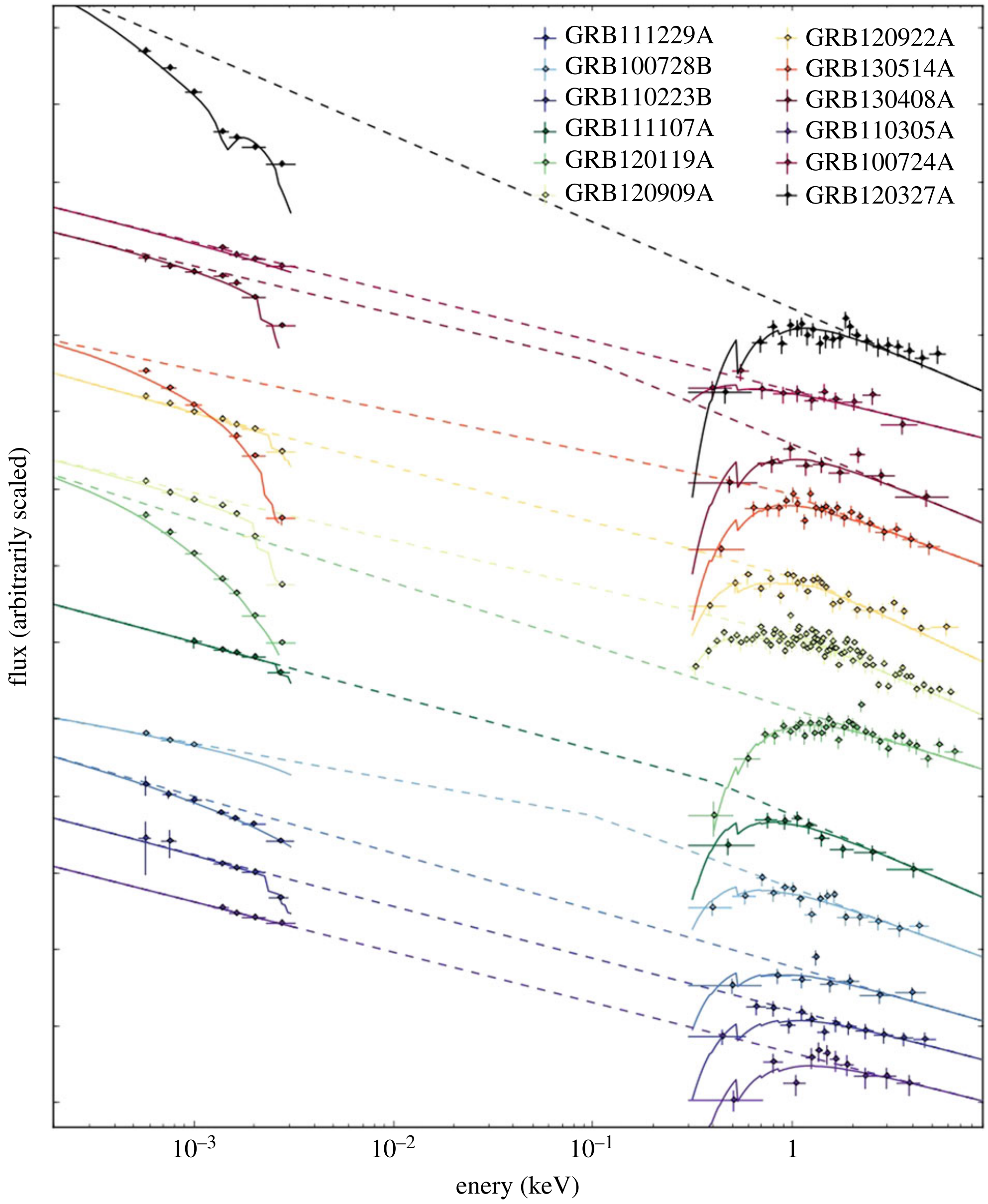

**Figure 8.** Observed GRB *Swift* and GROND afterglow SEDs (data points and solid lines). The best-fit intrinsic power law or broken power law spectral fits (dashed line) are also shown. Such an example GROND/*Swift* SED is routinely measured for approximately 40 GRBs per year [266].

The most commonly studied extinction curves are those of the Milky Way, and the Large (LMC) and Small Magellanic Clouds (SMC). The Milky Way extinction curve shows a very prominent dust extinction feature centred at approximately 2175 Å [264], which is weaker along LMC lines of sight and largely undetected within the SMC [265], and the steepness of the three extinction curves at UV wavelengths is anti-correlated with the prominence of the 2175 Å bump. GRBs offer the only effective way of studying dust extinction curves in a wider range of environments, in galaxies beyond the Local Group. Intervening systems along QSO lines of sight suffer from the fact that they are typically too dust-poor to leave a notable dust extinction imprint on the QSO spectrum, and they are generally undetected in emission, leaving the properties of the intervening absorption systems largely unknown.

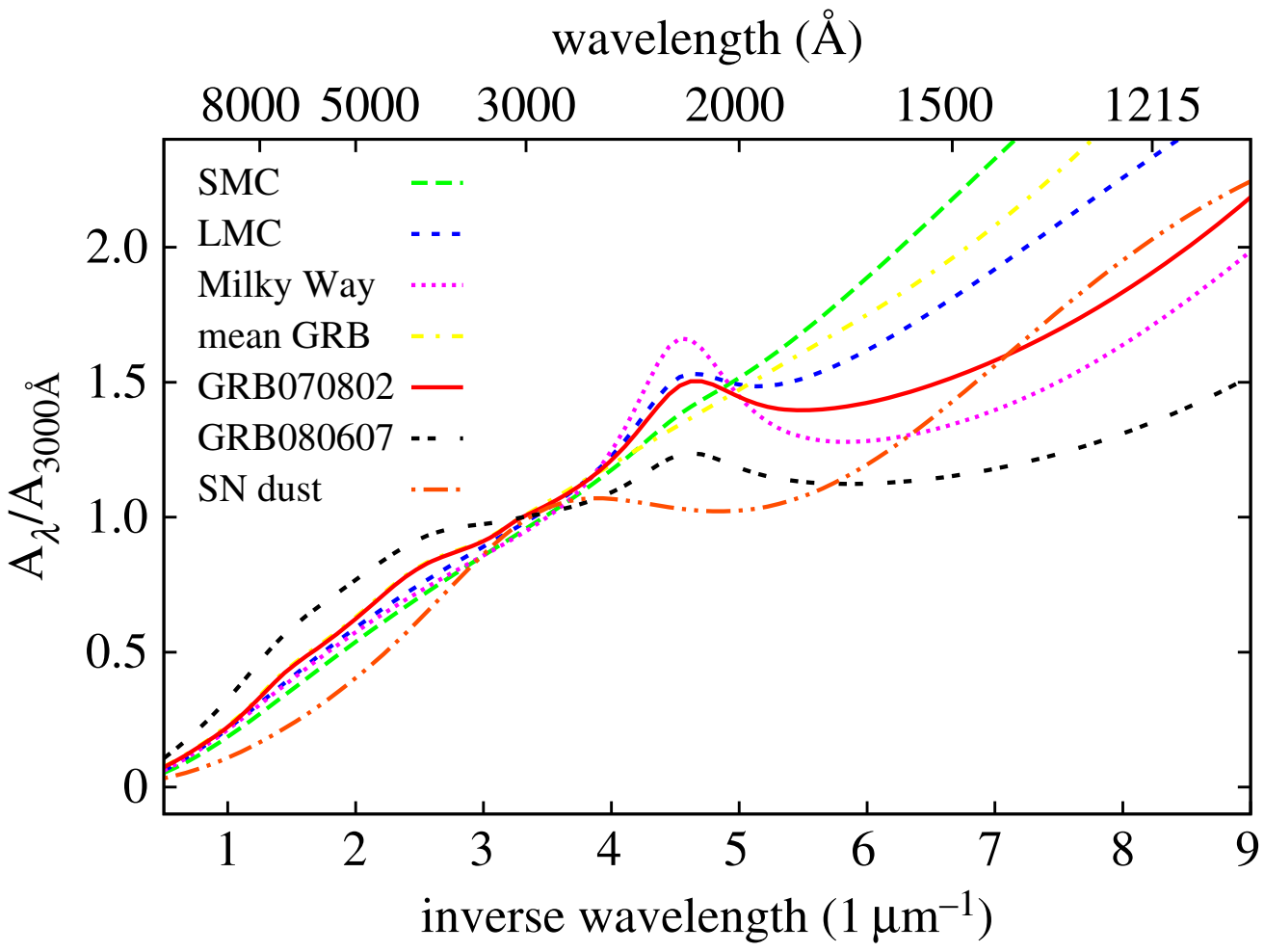

**Figure 9.** Average extinction curves from numerous lines of sight to the SMC (large-dash; green), the LMC (small-dash, blue), to the Milky Way (dots; pink), and to a sample of GRBs (dot-dash yellow; [233]). The figure also shows the extinction curves best-fit to the SED of GRB 070802 (solid red; [231]), GRB 080607 (blank-dash black; [275]) and the theoretical extinction law from SN-produced dust (dot-dot-dash orange; [276]).

Most analysis on GRB host galaxy extinction curves is done by modelling the broadband NIR-to-X-ray afterglow spectral energy distribution (SED), where the intrinsic afterglow spectral slope is well set by the largely unattenuated IR and hard X-ray wavelengths ($\gtrsim$2 keV) (figure 8). Such analysis constrains very well the overall shape of the extinction law, although particular features, such as the 2175 Å bump are not well resolved. For this reason, GRB afterglow SEDs are nominally fitted using template extinction curves taken from average Milky Way, SMC and LMC lines of sight, which generally fit the data very well. There are certainly exceptions to this, and claims have been made for extinction curves along GRB lines of sight that differ significantly from the Local Group, such as evidence for very flat extinction curves [267,268], which would be suggestive of a grain size distribution skewed to large grains, as may be expected in the event of significant dust destruction. GRB sight lines with broadband extinction reminiscent of what would be produced by dust formed through SNe explosions have also been reported [269,270], although such claims are controversial in terms of the significance of the detection [271], and on even the possibility of detecting such an 'SN extinction curve' [272].

The more common population of relatively unextinguished GRB lines of sight (typical visual extinction values of $A_V \sim 0.3$ mag) show little curvature in their SEDs and are well fitted with the relatively linear SMC-like extinction law. Some GRB sight lines with high signal-to-noise optical and NIR data have shown indication for flatter, Milky Way-like host extinction curves [230,273], but the first firm detection of a 2175 Å extinction feature in a GRB afterglow was not until GRB 070802, where the absorption imprint was clearly seen in photometric [274] and spectroscopic data [231]. This GRB also had one of the highest measured host galaxy visual extinctions at the time ($A_V = 0.8$–$1.8$ mag). The detection rate of GRBs with Milky Way or LMC-like host galaxy dust extinction curves along the line of sight has since increased, and now makes up 10–15% of sight lines with good, broadband afterglow data [135,232], most of which have large visual extinctions ($A_V \gtrsim 1.0$ mag; [135,192,232]). The increase in the detection rate of significantly dust-extinguished GRBs during the *Swift* era has thus greatly contributed to the rise in the detection of the 2175 Å dust-extinction feature. Additionally, there is some evidence that the most dust-extinguished GRBs have host galaxy extinction curves that are even flatter than observed in the Milky Way (figure 9).

Most claims of a detected 2175 Å extinction feature in GRB afterglows are based on photometric data, and although these detections are often highly significant, spectroscopic data are necessary to study the profile of the feature. In four examples where a 2175 Å dust-extinction feature was detected in the afterglow spectrum [231,237,277], the central wavelength and width of the extinction feature were consistent with what is observed in the Local Group, but the strength of the bump (depth of the feature) was weaker along GRB sight lines for the same given $A_V$ [231,277]. However, the fifth and most recent spectroscopic detection of a 2175 Å extinction feature had a profile that was much wider, and stronger than what is generally seen along lines of sight through the Milky Way [278]. Although GRB sight lines now make up the majority of firm detections of the 2175 Å bump extinction feature outside of the Local Group, the sample is still too small to carry out a comprehensive analysis on the origin of this feature. Interestingly, two of the GRB afterglows with detections of the 2175 Å bump also had absorption features from molecular gas [237,278], although it is likely that they are indirectly linked by a mutual propensity to arise in dusty environments. Highly dust extinguished yet bright GRBs, where a decent signal-to-noise afterglow spectrum is obtainable are rare. Nevertheless, the sample of individual extragalactic extinction curves is sufficiently small that every additional spectroscopic detection of the 2175 Å bump and well-measured extinction curves from GRB afterglows greatly enhance the sample.

## 4.3. Dust-to-metals ratio

Another complementary method to study the origin of dust and its evolution with redshift and environment is to analyse the dust-to-metals ratio. Metals are formed through nucleosynthesis within stars, and they are expelled into the ISM of the galaxy via strong stellar winds and SN explosions. A fraction of these metals can cool and condense into dust grains, and assuming that the efficiency of dust formation in stellar winds and SNe remains fairly constant with redshift, the dust-to-metals ratio arising from stellar processes alone will show little evolution. Another dust formation mechanism that is believed to be important is *in situ* grain growth within the ISM [279]. *In situ* dust formation does not affect the total metal budget, and it will thus cause the dust-to-metals ratio to increase if at any stage it begins to contribute significantly to the dust mass of a galaxy. Interestingly, all channels of dust production are seemingly too inefficient to account for the large dust masses inferred to have already been in place in some distant galaxies [280], and along the lines of sight to high-redshift QSOs (e.g. quasar J1148+5251 at $z = 6.4$; [281]).

The plethora of metal absorption lines imprinted on GRB afterglow spectra from the host galaxy ISM, in addition to the dust imprint, offers an opportunity to study the dust-to-metals ratio of extragalactic galaxies down to $10^{-2}$ solar metallicity and out to $z > 5$ [218,219,282]. Such measurements are also possible with QSO–DLA spectral observations [218,282,283], although the GRB observations have the advantage that they directly probe the star-forming regions of a galaxy, where metals and dust are more abundant. Since the DLAs frequently observed in GRB host galaxies (see §3.2) and along the line of sight to QSOs imply negligible ionization corrections, the observed absorption imprint from singly ionized metals is an accurate tracer of the amount of metals in the gas phase along the QSO or GRB line of sight. The column of dust along the same line of sight can be traced from either the amount of dust extinction that the background source has undergone (i.e. $A_V$; §4.2), or by the fraction of metals along the line of sight that are missing or *depleted* from the gas phase because they are locked onto dust grains. More refractory elements, such as iron and nickel, deplete more quickly onto dust grains than volatile elements such as zinc and silicon, and their relative abundances compared to solar can thus be used to derive a measure of what fraction of metals are missing from the gas phase, and thus how much dust there is along the line of sight. For a more detailed review on the relative abundances and dust depletion in GRB afterglows, I refer the reader to [284].

Using the visual extinction, $A_V$, measured from the GRB afterglow SED to trace the dust, no evidence for evolution in the dust-to-metals ratio was found by Zafar & Watson [282] covering over three orders of magnitude in metallicity, and out to $z \sim 5$. They took this as evidence that *in situ* dust formation is not an efficient mechanism. These results are in contrast with those of [218], where the dust column along GRB and QSO lines of sight from the depletion of Fe measured in optical spectra were derived and significant evolution in the dust-to-metals ratio with metallicity over a comparable metallicity range as in [282] was found. Such evolution in the dust-to-metals ratio was later confirmed by Wiseman *et al.* [219], who used a minimum of four singly ionized metal lines detected in the GRB spectrum to accurately measure the dust depletion along 19 GRB lines of sight. The results from [218,219] imply that *in situ* dust formation becomes increasingly efficient at higher metallicities. When [219] used the afterglow SED extinction, as in [282], rather than dust depletion to trace the dust on the same sample of GRB afterglows, they no longer found a strong trend between the dust-to-metals ratio with metallicity. It is therefore the method rather than differences in samples that produces the conflicting results on the metallicity dependence of a galaxy's dust-to-metals ratio. Wiseman *et al.* [219] largely ruled out systematic effects in the extinction and depletion measurements, with neither inaccurate SED model fits nor contamination from intervening systems providing natural explanations for the lack of a clear relation between dust extinction and depletion. The implication is that either dust extinction and depletion are very loosely related, or that they trace different populations of dust; neither of which seem very intuitive. A larger sample of well-observed dusty GRB sight lines should help understand the origin of this problem, as well as additional sight lines to QSO–DLAs. For now it seems fair to assume that the dust-to-metals ratios measured from dust depletion analysis is the more accurate, given that this technique derives the column of dust and metals simultaneously.

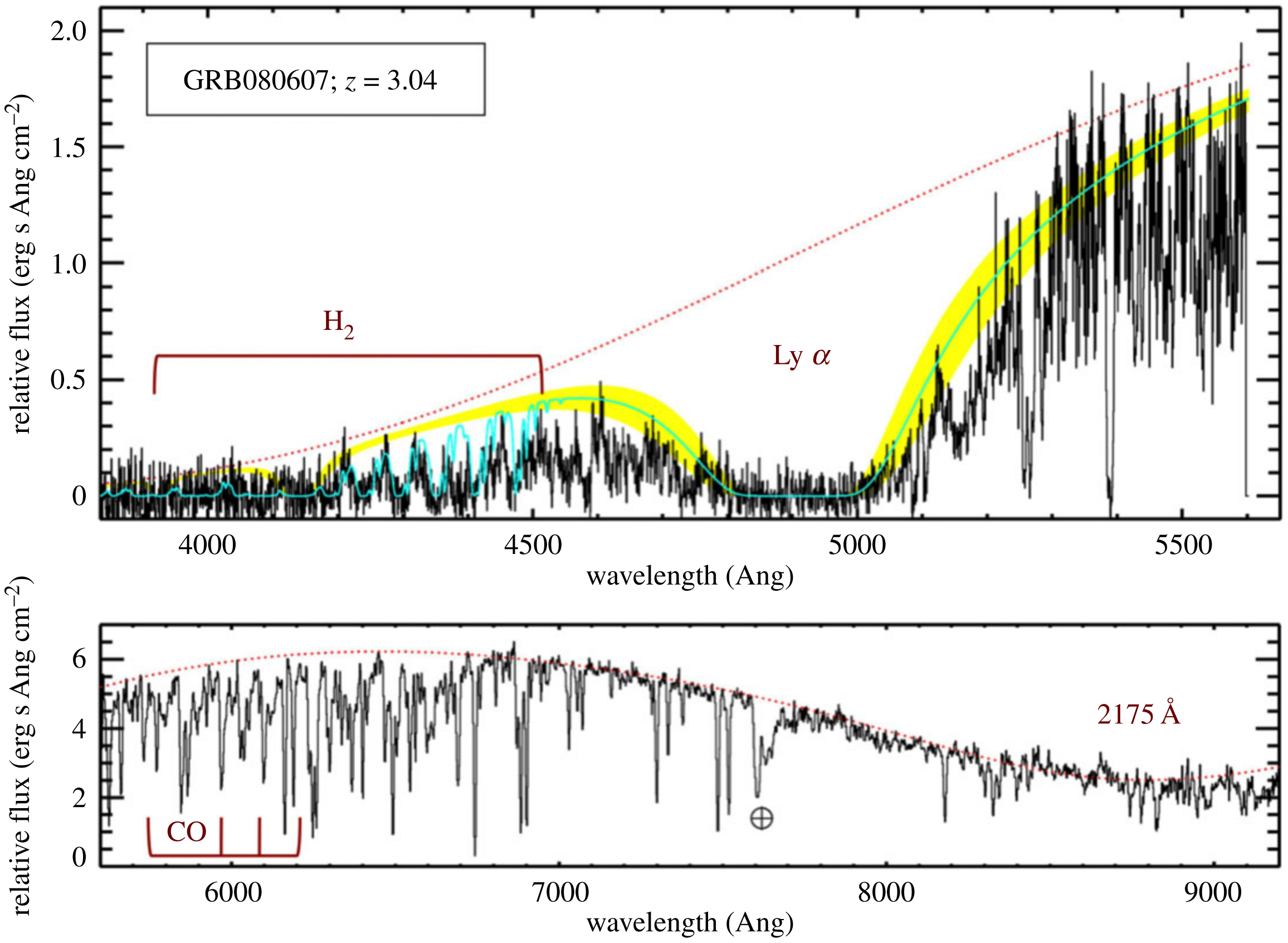

**Figure 10.** Observer-frame Keck/LRIS spectra of the optical afterglow of GRB 080607, with best-fit model of the intrinsic afterglow spectrum reddened by dust overplotted (dashed red) (adapted from [237]). Clearly seen is a Lyman-$\alpha$ absorption feature centred at 4900 Å, numerous metal absorption lines at a common redshift of $z = 3.0363 \pm 0.0003$, the first solid detection of $H_2$ and the only detection of CO to date. A broad absorption trough centred at 8800 Å is also present, resulting from host galaxy dust with a 2175 Å Milky Way-like extinction feature. The yellow shaded region corresponds to an H I column density $N_{\rm HI} = 10^{22.7\pm0.2}\ \rm cm^{-2}$. The cyan solid line indicates the best-fit model of the neutral hydrogen absorption within the host galaxy, including both atomic and molecular gas. Absorption from the Earth's A-band is marked by a $\oplus$. Reproduced from Prochaska *et al.* [237] (Copyright 2009 AAS).

## 4.4. The paucity of molecular gas

Further insight on the conditions of the star-forming regions probed by GRBs is gauged from absorption of the afterglow from molecular hydrogen. In view of the direct link between GRBs and massive stars and the tight correlation between molecular gas and star-formation surface density in galaxies (Kennicutt–Schmidt or KS law; [285,286]), the detection of $H_2$ absorption features in afterglow spectra of GRBs at $z > 2$ was expected to be a common occurrence during the *Swift* era. However, in contrast with expectations, $H_2$ has rarely been detected in absorption, which is all the more surprising when considering the very large column densities of atomic hydrogen generally observed (figure 6). More specifically, the average neutral hydrogen column densities measured from GRB afterglow spectra correspond to neutral hydrogen gas surface densities of approximately $30\,M_\odot\,\rm pc^{-2}$, whereas in spiral galaxies, all gas in excess of approximately $9\,M_\odot\,\rm pc^{-2}$ is in molecular form [287]. Perhaps, a more appropriate comparison is with the pencil beam sight lines to QSO–DLAs, where the detection rate of $H_2$ increases significantly in those QSO–DLA sight lines with $\log N_{\rm HI}\,\rm cm^{-2} > 21.5$ (approx. 70%; [288]). Such large H I column densities are a common feature along GRB lines of sight and yet $H_2$ absorption has rarely been detected. The first unambiguous detection of molecular hydrogen absorption imprinted on an afterglow spectrum was in the case of GRB 080607 at $z = 3.0363$ [237] (figure 10), over 3 years after the launch of *Swift*. Since then there have only been a further three unambiguous detections of $H_2$ absorption features in GRB optical afterglows [289–291].

High-energy emission from the GRB is only capable of photodissociating the molecular gas out to a few parsecs [220,292], and the few studies that have explored the reasons for this apparent dichotomy have concluded that far UV radiation fields 10–100 times the Galactic mean field suppress the formation of the molecules [220,293]. Furthermore, detailed analysis on the conditions within the GRB natal H II region implies that the lack of $H_2$ can be explained by low metallicities ([X/H] < −1), low depletion factors and, at most, moderate particle densities [294]. This is somewhat in conflict with the derived environmental properties inferred from observations, and modelling of different data. For example, most GRBs have metallicities [X/H] > −1 (from both absorption and emission measurements; [219,222]), whereas only four GRBs to date have a robust detection of $H_2$ absorption in their UV rest-frame afterglow spectra.

The implied conditions place strong constraints on the natal star-forming regions traced by GRBs. The very high far UV radiation fields inferred may be suggestive of a stellar mass distribution skewed to more massive stars (i.e. top heavy IMF). Alternatively, the apparent violation of the KS law in GRB host galaxies could reflect the age of the stellar population at the position of the GRB. Molecular hydrogen traces the very early stages of star formation, whereas more evolved stellar populations will have dissociated a larger fraction of the surrounding molecular gas [295–299]. The general paucity of molecular gas in GRB environments may thus be a reflection of the age of the stellar populations traced by GRBs and of their progenitors, which although massive, are not extremely young or massive at the time of explosion [209,300].

Until now there has not been any attempt to derive molecular gas column density limits on the currently large sample of high-quality afterglow spectra taken with *X-shooter*. This is necessary to quantify the selection effects biased against the detection of $H_2$ in GRB afterglow spectra, and to have a comprehensive view of the 'paucity' of molecular gas in GRB host galaxies.

## 5. Future prospects for gamma-ray burst research

*Swift* was launched at a time when GRB research was progressing at a fast rate, but where high-quality afterglow data were scarce. The rapid and accurate GRB afterglow positions it provides, and the many dedicated multi-band and spectroscopic follow-up campaigns during the past decade have signified a leap in our understanding of GRBs and the environments in which they are formed. Over 12 years after its launch, the *Swift* mission continues to observe GRBs, outliving its initial 2 years of funding sixfold. The Chinese/French Space-based multiband astronomical Variable Objects Monitor (SVOM [301]), due to be launched in 2021, will continue the *Swift* legacy, building on the discovery of the past decade by extending its optical coverage to redder wavelengths (400–950 nm), thus increasing its sensitivity to more distant and more dust-extinguished GRB afterglows.

Nevertheless, further ground-breaking advances in the field of GRB astronomy are likely to arise from the multidisciplinary observations of GRBs from space-borne and ground-based facilities. For example, recent developments in instrumentation to acquire rapid polarimetry data of the GRB prompt emission (e.g. POLAR; [302]), and an increase in the number of GRBs with early time optical afterglow polarimetry measurements [303–305] will provide new insight on the dominant emission mechanisms, and possibly some clues on how the jets are formed. The recent availability of rapid, target of opportunity observations with the very sensitive Atacama Large Millimeter/submillimetre Array (ALMA) offers the chance to acquire very broadband coverage of the GRB afterglow, which is necessary to break certain degeneracies and pin down some of the detailed physics within the shocked regions (e.g. the accelerated electrons' energy distribution, fraction of energy within magnetic fields and shocked electrons [137]).

In contemplating their use as probes of the high-redshift Universe, it is important to emphasize the need for sensitive, mid/high-resolution spectroscopic afterglow observations to confirm the redshift and to measure the attenuation from the intervening material, without which their potential as cosmic probes remains largely untapped. In recent years, the mid-resolution broadband *X-shooter* spectrograph on the VLT has been doing most of the legwork to obtain high-quality afterglow spectra of Southern hemisphere GRBs, but the mid- to high-resolution optical spectral coverage of northern GRBs has greatly waned in the past few years. Future state-of-the-art observatories, such as the forthcoming JWST (James Web Space Telescope) mission or the E-ELT (ESO Extremely Large Telescope) and other very large telescopes will offer new opportunities to study the high redshift Universe with GRBs. These observatories will provide very sensitive host galaxy observations, with the potential to study in detail the spatially resolved properties of their hosts. Spectral data on approximately 500 pc scales are currently only possible for the nearest 2% of long GRB host galaxies ($z \lesssim 0.1$) with sensitive IFUs such as MUSE. Future IFU instruments on JWST, for example, will provide equivalent spatial resolution for host galaxies out to $z \sim 0.3$. Ultimately, the aspiration would be to obtain spatially resolved spectroscopic data of the most heavily dust-obscured GRBs (generally at $z > 1.0$ [192,193]), and to then study the differences in the local environments of dust-enshrouded and dust-free GRBs.

Further in the future, the planned *Athena* X-ray mission promises to provide exquisite X-ray afterglow spectra of every GRB that it observes (around 40 per year), resolving the absorption lines from intervening high-ion metals. These data alone will provide GRB redshifts and a detailed description of the hot gas within the GRB host galaxy, which combined with optical afterglow spectra, will conclusively resolve the origin of the X-ray absorption excess (§4.1). However, *Athena* itself will require external GRB triggers, and it is not yet clear what GRB missions will still be orbiting by the expected launch date of *Athena* in 2028, with the next generation GRB mission, SVOM, having a planned lifetime that currently only takes it to 2024. If GRB triggers are available some time during the 5-year lifetime of *Athena*, then undoubtedly the available afterglow X-ray spectra will provide the closest look of the conditions within the natal GRB H II region [246].

We are currently entering an era of multi-messenger astronomy, and GRBs are predicted to produce a gravitational wave signal, high-energy neutrinos and radiation across the electromagnetic spectrum, from TeV energies down to the lowest radio frequencies. Despite their rarity, the investigation of GRBs and their use as cosmic probes will thus be an active area of research within a number of rapidly developing areas of astronomy. There is therefore much promise that over the coming decade we will continue to see notable developments in the field of GRB astronomy in the context of the emission mechanisms behind the explosion, the rates of short and long GRBs, their relation to star formation and the properties of their high-redshift galaxies.

Data accessibility. No new data have been generated for the purposes of this review article, and all data in this review are available from the sources referenced.
Competing interests. I have no competing interests.
Funding. Financial support came through the Sofja Kovalevskaja Award from the Alexander von Humboldt Foundation of Germany.
Acknowledgements. I dedicate this review to Neil Gehrels, who was not only the principal investigator of *Swift*, without which GRB afterglow research may well still be in its infancy, but also an important influence on the development of my own personal career and whom I have a lot to thank for. I also thank members of the GRB team at MPE for much fruitful discussion over the years that has helped develop and broaden my perspective on many of the scientific topics touched upon in this review, with particular emphasis on Thomas Krühler, Hendrik van Eerten, Sandra Savaglio, John Graham and Jochen Greiner.